\documentclass[twocolumn,english,pra,aps,showpacs]{revtex4-1}
\usepackage{amsmath}
\usepackage{amssymb}
\usepackage{graphicx}
\usepackage{esint}
\usepackage{color}
\usepackage{ulem}

\makeatletter

\@ifundefined{textcolor}{}
{%
\definecolor{BLACK}{gray}{0}
\definecolor{WHITE}{gray}{1}
\definecolor{RED}{rgb}{1,0,0}
\definecolor{GREEN}{rgb}{0,1,0}
\definecolor{BLUE}{rgb}{0,0,1}
\definecolor{CYAN}{cmyk}{1,0,0,0}
\definecolor{MAGENTA}{cmyk}{0,1,0,0}
\definecolor{YELLOW}{cmyk}{0,0,1,0}
}

\newcommand{\der}{\partial}

\@ifundefined{definecolor}

\makeatother

\usepackage{babel}
\begin{document}

\title{Two-body momentum correlations in a weakly interacting one-dimensional Bose gas}

\author{I. Bouchoule$^{(1)}$, M. Arzamasovs$^{(2)}$, K. V. Kheruntsyan$^{(3)}$  and D. M. Gangardt$^{(2)}$}
\affiliation{$^{(1)}$Laboratoire Charles Fabry, UMR8501 du CNRS, Institut d'Optique, 91127 Palaiseau, France\\
$^{(2)}$School of Physics and Astronomy, University of Birmingham, Edgbaston, Birmingham, B15 2TT, United Kingdom \\
$^{(3)}$The University of Queensland, School of Mathematics and Physics, Brisbane, Queensland 4072, Australia
}

\date{\today}

\begin{abstract}
  We analyze the two-body momentum correlation function for a uniform weakly
  interacting one-dimensional Bose gas. We show that the strong positive correlation
  between opposite momenta, expected in a Bose-Einstein condensate with a true
  long-range order, almost vanishes in a phase-fluctuating quasicondensate
  where the long-range order is destroyed. Using the Luttinger liquid
  approach, we derive an analytic expression for the momentum correlation
  function in the quasicondensate regime, showing (i) the reduction and
  broadening of the opposite-momentum correlations (compared to the singular
  behavior in a true condensate) and (ii) an emergence of
  \textit{anti}correlations at small momenta.  We also numerically
  investigate the momentum correlations in the crossover between the
  quasicondensate and the ideal Bose-gas regimes using a classical field
  approach and show how the anticorrelations gradually disappear in the ideal-gas limit.
\end{abstract}


\maketitle

\section{Introduction}

Many-body correlation functions contain valuable information about the physics
of quantum many-body systems and therefore their measurement constitutes an
important probe of the correlated phases of such systems. In recent years,
ultracold atom experiments have shown that atomic correlations can be
accessed via many experimental techniques,
including high-precision absorption \cite{Greiner2005,Folling2005,Esteve06} or fluorescence imaging 
\cite{Raizen:2005,Sherson2010,Bakr2010}, detection of atom transits through a high-finesse optical cavity \cite{Esslinger2005},
single-atom detection using multichannel plate detectors \cite{HBT,Jeltes:07,Hodgman2011,He-Review} or scanning electron microscopy techniques \cite{SEMdetection}, 
and the measurement of rates of two-body (photoassociation) \cite{Weiss:2005} or three-body loss processes \cite{Phillips-2004,Itah-3body,Whitlock:2010}.
While the loss-rate measurements depend only on local correlations, 
the imaging and atom detection techniques typically depend on nonlocal correlations which are embedded 
in the atom number fluctuations in small detection volumes (such as image pixels) or in the coincidence counts of time- and position-resolved atom detection events.

The development of these techniques have enabled the study of a wide range of phenomena in ultracold atomic gases, 
including the Hanbury Brown--Twiss effect \cite{HBT,Esslinger2005,Rom2006,Jeltes:07,Hodgman2011,Truscott:speckle,HBTPerrin} and higher-order coherences \cite{ANU-He}, phase fluctuations in quasicondensates \cite{Phase-fluctuating-Hannover,Manz2010}, superfluid to Mott insulator transition \cite{Folling2005,Spielman2007,Sherson2010,Bakr2010}, 
isothermal compressibility and magnetic susceptibility of Bose and Fermi gases \cite{Esteve06,ArmijoSkew,Sanner10,Mueller10,Sanner10,Sanner:11,Esslinger2012}, scale invariance of two-dimensional (2D) systems \cite{Hung2011}, the phase diagram of the 1D Bose gas \cite{Armijo10_2,Subbunching}, 
entanglement and spin squeezing in two-component and double-well systems 
\cite{Esteve2008,Oberthaler-interferometer,Riedel-interferometer,Hannover-twins,Chapman:11}, 
sub-Poissonian relative atom number statistics \cite{Jaskula2010,Vienna-twins,Hannover-twins}, and violation of the Cauchy-Schwarz inequality with matter waves \cite{Kheruntsyan-CS}.

From a broad statistical mechanics point of view, most of these measurements have so far given access to either equilibrium position-space 
density correlations or nonequilibrium momentum-space density correlations. In this paper, we address the question of equilibrium momentum-space density correlations \cite{Altman04} by focusing on the two-body correlation function
\begin{equation}
{\cal G}(k,k')=\langle\delta \hat{n}_{k}\delta \hat{n}_{k'}\rangle=\langle\hat{n}_{k}\hat{n}_{k'}\rangle - \langle\hat{n}_{k}\rangle \langle\hat{n}_{k'}\rangle,
\label{eq.Gkkp}
\end{equation}
for a weakly interacting uniform 1D Bose gas. 
Here, $\delta
\hat{n}_{k}=\hat{n}_{k}-\langle \hat{n}_{k}\rangle$ is the fluctuation in the population
$\hat{n}_{k}$ of the state of momentum $\hbar k$ [see
Eqs.~(\ref{eq:g1}),~(\ref{eq:g2}), and~(\ref{eq.Gkkpcorr})].

To measure ${\cal G}(k,k')$ experimentally, one needs to analyze atomic density fluctuations 
in a set of momentum distributions. Single-shot momentum distributions of a 1D
Bose gas, realizable by confining the atoms to highly anisotropic trapping
potentials, 
 can be acquired as follows. First, by turning off (or strongly
reducing) the transverse confinement, one ensures that atom-atom
interactions no longer play any role in the system dynamics. 
The longitudinal momentum distribution is 
unaffected by the turning off since, in 1D geometry, 
the turning-off time (which is on the order of the period of the
transverse confining potential) is much smaller than the relevant time scales
of the longitudinal (axial) motion of the atoms.
Second, the
longitudinal momentum distribution can, in principle, be measured using an
expansion along the long axis, 
after switching off the longitudinal confinement, or by 
using a recently demonstrated technique of Bose-gas
focusing~\cite{Shvarchuck2002,Amerongen08,momentumdistriJacqmin} (see also \cite{Davis:12,2Dfocusing}).

In the presence of a true long-range order, the Bogoliubov theory correctly
describes the excitations of a Bose condensed gas, predicting strong positive
correlations in ${\cal G}(k,k')$ between opposite momenta, $k^{\prime}=-k$,
for small $|k|$, as shown in Ref.~\cite{Mathey2009} (see also
\cite{Imambekov09}).  However, true long-range order is destroyed by
long-wavelength fluctuations in a 1D Bose gas~\cite{Hohenberg67}; for a large
enough system, the gas lies in the so-called quasicondensate regime
\cite{Petrov00} where, while the density fluctuations are suppressed as in a
true Bose-Einstein condensate (BEC), the phase still fluctuates along the cloud. In this paper, we show
that, when the system size becomes much larger than the phase correlation
length, the positive correlations between the opposite momenta vanish. In the
thermodynamic limit of an infinite quasicondensate, we find an analytic
expression for ${\cal G}(k,k')$ and show that it develops zones of
anticorrelation on the $(k,k^{\prime})$ plane.
We also
analyze the crossover from the quasicondensate to the ideal Bose gas regime,
using a classical field theory, and show how the behaviorof ${\cal G}(k,k')$
undergoes a continuous transformation between the two limiting regimes.

The paper is organized as follows. In Sec.~\ref{sec:generalities} we
outline the generalities applicable to two-body momentum correlations for the
uniform 1D Bose gas with contact interactions.  Section ~\ref{sec:true}
summarizes the known results in the regime of a true condensate. In Sec. \ref{sec:quasiBEC}, 
we show that the correlations between opposite momenta, which exist
in the case of a true BEC, disappear in the quasicondensate regime. Here we first use a
simple model of a quasicondensate (Sec.~ \ref{sec:simple}), followed by the
Luttinger liquid approach (Sec.~\ref{sec:Luttinger}) leading to an exact
analytic result for the two-body momentum correlation function. In Section
\ref{sec:crossover} we describe the momentum correlations in the crossover
from the quasicondensate up to the ideal Bose gas limit, using a classical
field method.  We discuss the experimentally relevant aspects in
Section~\ref{sec:experimental}, and conclude with a summary in Section~\ref{sec:summary}.

\section{Generalities}
\label{sec:generalities}

We consider a uniform gas of bosons interacting via a pairwise
$\delta$-function potential in a 1D box of length $L$ with periodic boundary
conditions. In the second-quantized form, the
Hamiltonian density is
\begin{eqnarray}
 \label{eq:ham}
 {\cal H}=  -\frac{\hbar^2}{2m} \hat{\psi}^\dagger
   \frac{\der^2}{\der z^2}\hat{\psi} + 
   \frac{g}{2}\hat{\psi}^\dagger\hat{\psi}^\dagger\hat{\psi}\hat{\psi}
   -\mu\hat{\psi^\dagger}\hat{\psi},
\label{eq.HLL}
\end{eqnarray}
where $\hat{\psi}(z)$ and $\hat{\psi}^\dagger(z)$ are the bosonic field
operators, $m$ is the mass of the particles, $g$ is the interaction
constant, and $\mu$ is the chemical potential. In the grand-canonical 
formalism that we are using, the equilibrium density
$\rho=\langle\psi^\dagger\psi\rangle$ is fixed by $\mu$ and the 
temperature $T$,
and the total number of particles is given by $N=\rho L$. Throughout this paper, we
restrict ourselves to the weakly interacting regime, which corresponds to the
dimensionless interaction parameter $\gamma=mg/\hbar^2\rho\ll 1$.

The momentum distribution $\langle \hat{n}_k\rangle$ and its correlation function
$\mathcal{G} (k,k')$ are related to the first- and second-order correlation functions of the bosonic
fields,
\begin{equation}
 \label{eq:g1}
 G_1(z_1,z_2)=G_1(z_1-z_2)=\langle \hat{\psi}^\dagger(z_1)\hat{\psi}(z_2)\rangle,
 \end{equation}
 and 
 \begin{equation}
 \label{eq:g2}
 G_2 (z_1,z_2,z_3,z_4) = 
 \langle \hat{\psi}^\dagger(z_1)\hat{\psi}(z_2)\hat{\psi}^\dagger(z_3)\hat{\psi}(z_4)\rangle,
 \end{equation}
 via the Fourier transforms
\begin{eqnarray}
\label{eq:nk}
\langle \hat{n}_k\rangle =\frac{1}{L} \iint_{0}^{L} \!dz_1 dz_2\;
 e^{-ik(z_1-z_2)} G_1(z_1,z_2), 
\end{eqnarray}
and 
\begin{eqnarray}
&&\mathcal G(k,k')=\frac{1}{L^2}\iiiint_{0}^{L}\!d^4 z\; e^{-ik(z_1-z_2)}
e^{-ik'(z_3-z_4)}  \nonumber\\
&& \times \left[ G_2(z_1,z_2,z_3,z_4)-G_1(z_1,z_2) G_1(z_3,z_4)\right]\, ,
\label{eq.Gkkpcorr}
\end{eqnarray}
where $d^{4}z\equiv dz_1 dz_2 dz_3 dz_4$. In Eq. (\ref{eq:g1}), the dependence 
of $G_1(z_1,z_2)$ only on the relative coordinate $z_1-z_2$ follows from the translational 
invariance of the system.

Several general statements about the momentum correlation function $\mathcal G(k,k')$ 
can be made, which are valid in any regime of the gas.
First, the correlation function obeys the following sum rule:
\begin{equation}
\sum_{k,k'}\mathcal G(k,k')=\langle \hat{N}^2\rangle - \langle \hat{N}\rangle^2,
\label{eq.sumrule}
\end{equation}
where $\hat{N}=\int_{0}^{L}dx \hat{\psi}^{\dag}(x)\hat{\psi}(x)$ is the total particle number operator. This implies that,
within the canonical ensemble, one has $\sum_{k,k'}\mathcal G(k,k')=0$. In the
grand canonical ensemble, the fluctuation-dissipation theorem, which connects the
particle number variance with the derivative of $\langle \hat{N}\rangle$ with
respect to the chemical potential $\mu$ \cite{Esteve06}, gives
\begin{equation}
\sum_{k,k'}\mathcal G(k,k')= k_BT \frac{\partial N}{\partial \mu}=k_BTL \frac{\partial \rho}{\partial \mu}.
\label{eq.sumrule2}
\end{equation}

Second, $\mathcal G(k,k')$ possesses several symmetries.  In thermal
equilibrium, the position-space correlation functions are invariant by the
simultaneous refection symmetry of all coordinates $z_i\rightarrow -z_i$. This
symmetry and the bosonic commutation relations between the field operators
imply, for periodic boundary conditions, that $\mathcal G(k,k')$ is symmetric
around the axis $k'=k$ and around the axis $k'=-k$.

Finally, for systems that have correlation lengths  
much smaller than the system size $L$, the two-body momentum correlation function 
$\mathcal G(k,k')$ can be
split into a `singular' part and a regular function.  
(We use the term `singular' in the sense of the Kronecker $\delta$-function, which turns into the Driac $\delta$-function singularity in the thermodynamic limit of $L\rightarrow \infty$.)
To show this, let us first
note that if we assume the existence of a finite correlation length $l_\phi$
for the decay of the first-order correlation function $G_1(z_1,z_2)$, then the
second-order correlation function $G_2(z_1,z_2,z_3,z_4)$ must have the
following two asymptotic limits:
\begin{equation}
\left \{
\renewcommand{\arraystretch}{1.5}
\begin{array}{l}
G_2(z_1,z_2,z_3,z_4)\simeq G_1 (z_1-z_2)G_1(z_3-z_4),\\
\mbox{for } 
|z_1-z_3|\gg l_\phi \mbox{ and } |z_1-z_2|,|z_3-z_4| \lesssim l_\phi, \\
\end{array}
\right.
\label{limit1}
\end{equation}
and 
\begin{equation}
\left \{
\renewcommand{\arraystretch}{1.5}
\begin{array}{l}
G_2(z_{1},z_{2},z_{3},z_{4})\simeq G_1(z_{1}-z_{4})G_1(z_{2}-z_{3})\\
\qquad\qquad\qquad\quad+G_1(z_{1}-z_{4})\delta(z_{2}-z_{3}),\\
\mbox{for }|z_{1}-z_{2}|\gg l_{\phi}\mbox{ and }|z_{1}-z_{4}|,|z_{2}-z_{3}|\lesssim l_{\phi}.
\end{array}
\right .
\label{limit2}
\end{equation}
In Eq.~(\ref{limit2}), the $\delta$-function term appears simply as a result
of normal ordering of the operators in Eq.~(\ref{eq:g2}).
By separating out the two asymptotic limits, Eqs.~(\ref{limit1}) and~(\ref{limit2}), 
we can write
\begin{eqnarray}
 \label{eq:g2free}
G_2 (z_1,z_2,z_3,z_4) &=& G_1 (z_1-z_2)G_1 (z_3-z_4)\nonumber\\
 &+&G_1(z_1-z_4)G_1 (z_2-z_3)\nonumber \\
 &+&G_1(z_1-z_4)\delta(z_2-z_3)\nonumber \\
 &+&\widetilde{G}_{2}(z_1,z_2,z_3,z_4),\
\end{eqnarray}
where $\widetilde{G}_{2}(z_1,z_2,z_3,z_4)$ is the remainder term.

By substituting Eq.~(\ref{eq:g2free}) into Eq.~(\ref{eq.Gkkpcorr}) we obtain 
\begin{eqnarray}
\mathcal G(k,k')= (\langle \hat{n}_k\rangle + \langle \hat{n}_k\rangle^2) \delta_{k,k'}
+ \widetilde{\mathcal G}(k,k'), 
\label{eq.GvsGtilde}
\end{eqnarray}
which shows explicitly that $\mathcal G(k,k')$ can be written down as a 
sum of a singular and regular contributions.
The first term in Eq.~(\ref{eq.GvsGtilde}) is the shot noise, the second term
is the bosonic ``bunching'' term, which describes the
exchange interaction due to Bose quantum statistics, and the last, regular term
$\widetilde{\mathcal G}(k,k')$ [the Fourier
transform of $\widetilde{G}_{2}(z_1,z_2,z_3,z_4)]$ describes the exchange of momenta between the 
particles 
during the binary elastic scattering processes and is nonzero
only for an interacting gas. For noninteracting bosons, Wick's theorem 
can be applied directly to the $G_2(z_1,z_2,z_3,z_4)$ function, which 
then leads to a vanishing $\widetilde{G}_{2}(z_1,z_2,z_3,z_4)$ and 
hence only to the singular terms in Eq.~(\ref{eq.GvsGtilde}).

\section{True condensate}
 \label{sec:true}

 At $T=0$, the first-order correlation function  
decays algebraically as  $G_1(z_1,z_2)\simeq (\xi/|z_1-z_2|)^{\sqrt{\gamma}/2\pi}$ for $|z_1-z_2|\gg \xi$
 \cite{Popov1980,Mora03,Cazalilla2004}, where $\xi=\hbar/\sqrt{mg\rho}$ is the
 healing length. As $\gamma \ll 1$ in the weakly interacting regime, 
the algebraic decay is very slow, leading to an exponentially large phase
correlation length. Indeed, defining  $l_{\phi}^{(0)}$ as the
length for which $G_1(z_1,z_2)$ decreases by a factor of $e$, 
we find  \cite{Petrov00}
\begin{eqnarray}
 \label{eq:lphi0}
 l_{\phi}^{(0)} \sim  \xi e^{2\pi/\sqrt{\gamma}}.
\end{eqnarray}

At finite temperatures, the algebraic decay of $G_1(z_1,z_2)$ remains valid for distances 
$\xi \ll |z|\ll l_{T}$, where $l_{T}=\hbar^2/mk_BT\xi=(\hbar/k_BT)\sqrt{g\rho/m}$ is the phonon thermal wavelength  \cite{Cazalilla2004,Mora03}. For distances $|z|\gg l_{T}$, on the other hand, the correlation function decays exponentially [see Eq. (\ref{eq.g1quasicond}) below] with the characteristic temperature-dependent phase coherence length 
\begin{equation}
l_\phi(T) = \hbar^2\rho/mk_BT.
\label{eq.lphi}
\end{equation} 
Considering now a 
system of size $L\ll \min\{l_{\phi}^{(0)},l_{\phi}\}$, we 
can assume true long-range order in the system and use the Bogoliubov theory 
to describe the momentum
correlations as was done in Refs.~\cite{Mathey2009,Imambekov09}. We briefly
recall the relevant results here.

The momentum correlation function ${\cal G}(k,k')$ is different from zero only
for $k=k'$ and $k=\!-k'$. For equal momenta $k=k'$, one finds ${\cal
  G}(k,k)=\langle \hat{n}_{k}\rangle+\langle \hat{n}_{k}\rangle^{2}$, which
is similar to the ideal Bose gas behaviour, except that the standard Bose
occupation numbers $\langle \hat{n}_{k}\rangle=(e^{(E_k-\mu)/k_BT}-1)^{-1}$
are now replaced by
\begin{eqnarray}
\label{eq:nkbec}
 \langle \hat{n}_{k}\rangle=
 (1+2\widetilde{n}_{k})\frac{E_{k}+g\rho}{2\epsilon_{k}}-\frac{1}{2}.
\end{eqnarray}
Here $\epsilon_{k}=\sqrt{E_k(E_k+2g\rho)}$ is the energy of the Bogoliubov modes,
$E_k =\hbar^{2}k^{2}/2m$ is the free particle dispersion, and
$\widetilde{n}_{k}=\left(e^{\epsilon_{k}/k_BT}-1\right)^{-1}$ are 
the mean occupation numbers of Bogoliubov
modes. For opposite momenta, $k=\!-k'$, one has
\begin{eqnarray}
 \label{eq:gk-k}
 {\cal G}(k,-k)=(1+2\widetilde{n}_{k})^{2}\left(\frac{g\rho}{2\epsilon_{k}}\right)^{2}.
\end{eqnarray}

A convenient way to characterize the relative strength of the opposite and
equal momentum correlations is via the normalized pair correlation function
\begin{equation}
{\cal P}(k)=\frac{{\cal G}(k,-k)}{{\cal G}(k,k)} = 1-\frac{\langle(\hat{n}_{k}-\hat{n}_{-k})^{2}\rangle}
{2\langle\delta \hat{n}_{k}^{2}\rangle}.
\end{equation}
Here, \mbox{${\cal P}(k)=1$} corresponds to perfect (maximum) correlation 
between the opposite momenta, whereas ${\cal P}(k)=0$ corresponds to the absence of any correlation.

At $T=0$, one has $\widetilde{n}_k=0$ and ${\cal G}(k,-k)={\cal G}(k,k)$, and therefore the Bogoliubov theory 
predicts perfect correlation between the opposite momenta, ${\cal P}(k)=1$. 
Such perfect correlation stems from the fact that the depletion of the
condensate in the Bogoliubov vacuum simply corresponds to the creation of pairs of
particles with equal but opposite momenta.

At finite temperatures, $\widetilde{n}_{k}$ is different from zero, nevertheless
the normalized pair correlation is still close to its maximum (perfect correlation) value,
${\cal P}(k)\simeq 1$, for phonon excitations with $k\ll 1/\xi$ 
for any value of $\widetilde{n}_{k}$. This 
can be understood from the fact that the phonons are mainly phase fluctuations,
so that they correspond to equal-weighted sidebands at momenta $k$ and $-k$ of the excitation spectrum. 
On the other hand, for particlelike excitations, with $k\gg 1/\xi$, the
thermal population of particles leads to a decrease of ${\cal P}(k)$.
More precisely, for $1/\xi<k < \sqrt{mk_BT}/\hbar$,
which corresponds to particle-like excitations whose occupation numbers are large,
one obtains ${\cal P}(k)\ll 1$. Finally, at very large momenta, $k \gg \sqrt{mk_BT}/\hbar$, 
for which the occupation numbers are negligibly small, one 
again recovers the zero-temperature result
${\cal P}(k)\simeq 1$.

\section{Quasicondensate regime}
\label{sec:quasiBEC}

\subsection{Effect of phase fluctuations}
\label{sec:simple}

The above results obtained using the Bogoliubov theory 
are valid  when the temperature is small
enough 
so that the phase correlation length is much larger than the system size,
$l_{\phi}\gg L$.
While this condition is easier to satisfy in 3D or quasi-1D systems,
it is generally not 
fulfilled for purely 1D gases.

In a large enough 1D system or at high enough temperatures, the long-range order 
is destroyed by long-wavelength 
phase fluctuations, having a characteristic temperature-dependent correlation 
length $\l_{\phi}$. When $l_{\phi}\ll L$, such a system is said to enter into the so-called 
quasicondensate regime \cite{Petrov00},
in which the density fluctuations are suppressed while the phase still fluctuates.
As we show here, the two-body correlation between opposite
momenta is expected to vanish in the quasicondensate regime.

To give a crude, yet simple estimate of the two-body momentum correlations,
we can divide the system into domains of length $l_{\phi}$ and 
assume that (i) within each domain,
the spatial variation of the phase is small,
and therefore the Bogoliubov approach 
for a true condensate can be applied to each domain, and (ii) the relative phases 
between two different domains are uncorrelated. 
For each domain, indexed by $\alpha$, the field operator $\hat{\psi}_{\alpha}(z)$
can be expanded according 
to the Bogoliubov theory, 
\begin{equation}
\hat{\psi}_{\alpha}(z)=e^{i\phi_{\alpha}}\left(\sqrt{\rho}+\frac{1}{\sqrt{l_{\phi}}}\sum_{k\neq0}
\delta\hat{\psi}_{\alpha,k}\,e^{-ikz}\right),
\label{eq.bogozone}
\end{equation}
where the first term is the mean-field component, the second term 
 is the fluctuating component expanded in terms of plane-wave 
 momentum modes $\delta\hat{\psi}_{\alpha,k}$, $\phi_{\alpha}$ is the mean global phase of the domain 
assumed to be a random variable distributed uniformly between $0$ and $2\pi$, and
the summation is over the momenta that are quantized in units of
$2\pi/l_\phi$.

Using the fact that the momentum component $\hat{\psi}_{k}=\frac{1}{\sqrt{L}}\int_{0}^{L}dz\hat{\psi}(z)e^{ikz}$ of the full field $\hat{\psi}(z)$ 
can be decomposed as $\hat{\psi}_{k}=\sqrt{l_{\phi}/L}\sum_{\alpha}\delta\hat{\psi}_{\alpha,k}e^{i\phi_{\alpha}}$ for $k\neq 0$, we obtain the following expression for the momentum correlation function:
\begin{align}
\label{eq:fourcorr}
\langle \hat{n}_{k}\hat{n}_{k'}\rangle &=\left(\frac{l_\phi}{L}\right)^2
\sum_{\alpha\beta\gamma\delta}\langle
\delta\hat{\psi}^\dagger_{\alpha,k}
\delta\hat{\psi}^{\phantom{\dagger}}_{\beta,k}
\delta\hat{\psi}^\dagger_{\gamma,k'}
\delta\hat{\psi}^{\phantom{\dagger}}_{\delta,k'}
\rangle \nonumber \\
& \times \overline{e^{-i(\phi_{\alpha}-\phi_{\beta}+\phi_{\gamma}-\phi_{\delta})}}, 
\quad (k,k^{\prime}\neq 0).
\end{align}
Here, the overline above the exponential factor
stands for averaging over the random mean phases of different domains.

Within the Bogoliubov theory, the Hamiltonian is quadratic 
in $\delta\hat{\psi}_{\alpha,k}$ and one can use
Wick's theorem to evaluate the four-operator correlation function in
Eq.~(\ref{eq:fourcorr}). Only pairs of operators belonging to the same domain
give a nonzero contribution since different domains are uncorrelated.
Among these pairs, only terms 
$\langle\delta\hat{\psi}_{\alpha,k}^\dagger\delta \hat{\psi}^{\phantom{a}}_{\alpha,k}\rangle$,
$\langle\delta\hat{\psi}_{\alpha,k}\delta\hat{\psi}_{\alpha,-k}\rangle$, and
$\langle\delta\hat{\psi}_{\alpha,k}^\dagger\delta\hat{\psi}_{\alpha,-k}^\dagger\rangle$
survive.

To evaluate these terms in the most transparent way we make use of the 
classical field approximation \cite{Castinatomlaser} 
(see also Sec. \ref{sec:classfield}), treating
the operators $\delta\hat{\psi}_{\alpha,k}$ and $\delta\hat{\psi}^{\dag}_{\alpha,k}$ as 
$c$-numbers, $\delta\psi_{\alpha,k}$ and $\delta\psi^{*}_{\alpha,k}$, and assuming that the respective 
mode occupations 
$\langle\hat{n}_{k}\rangle=\langle\delta\psi_{\alpha,k}^{*}
\delta\psi^{\phantom*}_{\alpha,k}\rangle$ are much larger than one, $\langle
\hat{n}_{k}\rangle\!\gg \!1$. For $k\!\ll \!1/\xi$, the excitations in each domain are almost
purely phase fluctuations so that
$\delta\psi_{{\alpha,-k}}=-\delta\psi_{\alpha,k}^{*}$ and therefore
$\langle\delta\psi_{\alpha,k}^{*}\delta\psi_{\alpha,-k}^{*}\rangle\simeq-\langle
\hat{n}_{k}\rangle$ \cite{c_field_and_Bog}. 
As a result, for the regular part of the momentum correlation function
we obtain
\begin{equation}
\widetilde{\cal G}(k,k')\simeq
\delta_{k,-k'}\left(\frac{l_{\phi}}{L}\right)^{2}
\sum_{\alpha,\beta} \overline{e^{-2i(\phi_{\alpha}-\phi_{\beta})}}\langle \hat{n}_{k}\rangle^{2}.
\label{eq.corrhw}
\end{equation}
Averaging 
over the phases gives $\overline{e^{-2i(\phi_{\alpha}-\phi_{\beta})}}=\delta_{\alpha,\beta}$, which
singles out only the diagonal in $\alpha$ and $\beta$ terms; there are
$L/l_\phi$ such terms in the sum in Eq.~(\ref{eq.corrhw}). Accordingly, for the correlation function 
with opposite momenta, we find
${\cal G}(k,-k)\simeq(l_{\phi}/L)\langle n_{k}\rangle^{2}$, 
whereas the correlation function for equal momenta is given by
${\cal G}(k,k)=\langle n_{k}\rangle^{2}+\langle n_{k}\rangle\simeq 
\langle n_{k}\rangle^{2}$, for $\langle n_{k}\rangle \gg 1$.
Therefore, for the 
normalized pair correlation ${\cal P}(k)$ we obtain the following simple result
\begin{equation}
\label{eq:pkscaling}
{\cal P}(k)\underset{k\ll1/\xi}{\simeq} \frac{l_{\phi}}{L} \ll 1,
\end{equation}
which shows that the correlations between the opposite momenta 
are inversely proportional to the system size $L$
and therefore are vanishingly small for $L\gg l_{\phi}$.

The above simple model is not capable of capturing features of ${\cal
  \widetilde{G}}(k,k^{\prime})$ on momentum scales smaller than, or of the order of,
the inverse phase correlation length, $k\lesssim 1/l_{\phi}$. 
For such momenta, the 
two-body correlation function is calculated below using a more
rigorous Luttinger liquid approach. The results obtained within this approach confirm
the simple scaling behaviorobtained in Eq. (\ref{eq:pkscaling}).  
Moreover, the Luttinger liquid results show that the
correlation function between different momenta is no longer singular on the
antidiagonal $k'=-k$ and that it develops zones of \textit{anti}correlation.

\subsection{Two-body correlations in the Luttinger liquid approach}
\label{sec:Luttinger}

The condition for the quasicondensate regime \cite{KheruntsyanPRL03} is 
\begin{equation}
T\ll T_{\rm{co}}\equiv \sqrt{\gamma}\frac{\hbar^2 \rho^2}{2mk_B}.
\label{eq.Tco}
\end{equation}
In this regime, the correlation functions in Eqs.~(\ref{eq:g1})
and~(\ref{eq:g2}) are dominated by the long-wavelength (low-energy)
excitations and the Hamiltonian density reduces to that of the Luttinger liquid 
\cite{Haldane:81,Cazalilla2004}:
\begin{eqnarray}
 \label{eq:HLutt}
 \mathcal{H}_L = \frac{g}{2} (\delta\hat{\rho})^2 +
   \frac{\hbar^2\rho}{2m}(\partial_z\hat{\phi})^2\, .
\end{eqnarray}
Here, $\delta\hat{\rho}(z)$ is the operator describing the density
fluctuations, canonically conjugate to the phase operator $\hat{\phi}(z)$,
with the commutator $[\delta\hat{\rho} (z), \hat{\phi}(z^{\prime})]
=i\delta(z-z^{\prime})$.

The density fluctuations are small in the quasicondensate regime
and, as long as the relative distances considered are much larger than the healing length $\xi$, they 
can be neglected when calculating the correlation functions (\ref{eq:g1}) and (\ref{eq:g2}) 
\cite{Mora03,Cazalilla2004,Sykes:08,Deuar:09}.
As the Luttinger liquid Hamiltonian given by Eq.~(\ref{eq:HLutt}) is
quadratic in $\hat{\phi}$, the first-order correlation function 
$G_1(z_1,z_2) =\rho\langle e^{i(\hat{\phi}(z_1)-\hat{\phi}(z_2))}\rangle $
can be expressed through the mean-square fluctuations of the phase using Wick's theorem:
\begin{eqnarray}
 \label{eq:g1wick}
 G_1(z_1,z_2) =\rho e^{-\frac{1}{2}\langle(\hat{\phi}(z_1)-\hat{\phi}(z_2))^{2}\rangle }.
\end{eqnarray}

Neglecting the contribution of vacuum fluctuations compared to thermal ones 
\cite{quantum_fluctuations} the calculation of the mean-square 
phase fluctuations leads to an exponentially decaying first-order correlation 
function \cite{Mora03,Cazalilla2004,Deuar:09},
\begin{eqnarray}
\label{eq:g1exp}
G_1 (z_1,z_2) = \rho e^{-|z_1-z_2|/2 l_\phi}, \quad(|z_1-z_2|\gg \xi).
\label{eq.g1quasicond}
\end{eqnarray}
This defines the finite-temperature phase coherence length $l_{\phi}$, given by Eq. (\ref{eq.lphi}),
and leads to a Lorentzian distribution for the momentum mode occupation numbers, 
\begin{equation}
\langle\hat{n}_k\rangle=\frac{4 \rho l_{\phi}}{1+(2l_{\phi}k)^2},
\label{eq.Lorentzian}
\end{equation} 
valid for $k\ll 1/\xi$.

Similarly, the two-body correlation function 	
$G_2(z_1,z_2,z_3,z_4)=\rho^2\langle
e^{i[\hat{\phi}(z_{1})-\hat{\phi}(z_{2})+\hat{\phi}(z_{3})-\hat{\phi}(z_{4})]}\rangle$
 can be represented in terms of the first-order correlation as
\begin{eqnarray}
 \label{eq.Cft}
&&G_2(z_1,z_2,z_3,z_4)\nonumber\\  
&&=\frac{G_1(z_1\!-\!z_2)G_1(z_3\!-\!z_4)G_1(z_1\!-\!z_4)
G_1(z_2\!-\!z_3)}
{G_1(z_1\!-\!z_3)G_1(z_2\!-\!z_4)}. 
\label{eq.g2quasicond} 
\end{eqnarray}

By substituting  Eqs.~(\ref{eq.g1quasicond}) and (\ref{eq.g2quasicond}) into
Eq.~(\ref{eq.Gkkpcorr}), we find that
the two-body momentum correlation function 
indeed has the form of Eq.~(\ref{eq.GvsGtilde}) \cite{nk}, 
in which the regular part 
$\widetilde{\cal{G}}(k,k')$ can be written as
\begin{equation}
\widetilde{\cal G}(k,k') = 
\frac{l_\phi}{L}\:
(\rho l_\phi)^2\: {\cal F}(2l_\phi k,2l_\phi k') \, ,
\label{eq.Fronde}
\end{equation}
where ${\cal F}(q,q^{\prime})$ is a dimensionless function given by
\begin{gather}
\mathcal{F}(q,q')= \frac{256}{(q^2+1)^2(q'^2+1)^2[(q+q')^2+16]}\nonumber \\
\times \left[(q^2+3qq'+q'^2)qq' -2 (q^2-qq'+q'^2)-7 \right]\, ,
\label{eq:fqbec}
\end{gather}
with $q\equiv 2l_{\phi}k$ and $q'\equiv 2l_{\phi}k'$. We note that the restriction of these 
results to $k\ll 1/\xi$ implies $q\ll l_{\phi}/\xi=2T_{\rm{co}}/T$, and 
that the scaling of $\widetilde{\cal G}(k,k')$
with the inverse size of the system $L$ coincides with the one 
obtained in Eq.~(\ref{eq:pkscaling}).

We now wish to check the constraints on 
the function $\widetilde{\cal G}(k,k')$ imposed by the sum rule, given by Eq.~(\ref{eq.sumrule2}). 
In evaluating the different terms 
in the left and right hand sides of 
Eq.~(\ref{eq.sumrule2}), we note that (i) for the derivative term we can use the equation of state for the quasicondensate regime, $\rho=\mu/g$,  (ii) the term $\sum _k \langle \hat{n}_k\rangle$ [coming from the singular part of ${\cal G}(k,k')$] is given simply by $\sum _k \langle \hat{n}_k\rangle=N=\rho L$, and (iii) the term $\sum _k \langle \hat{n}_k\rangle ^2$ can be evaluated 
using Plancherel's theorem and Eq.~(\ref{eq.g1quasicond}). As a result, the sum rule is reduced to \cite{limits}
\begin{equation}
\label{eq:sumrule1}
\frac{1}{(2\pi)^2} \iint_{-\infty}^{\infty} dqdq' \mathcal{F}(q,q') \simeq
-8+\left (\frac{T}{T_{\rm{co}}}\right )^2,
\end{equation}
where the contribution of the $\sum _k \langle \hat{n}_k\rangle$ is ignored on the grounds that it is of the order of $T/T_d$ (where $T_d=\hbar^2\rho^2/2mk_B$), which is always much smaller than unity in the  entire range of temperatures $T\lesssim T_{\rm{co}}$.

\begin{figure}[tbp]
~~~\includegraphics[width=7.00cm]{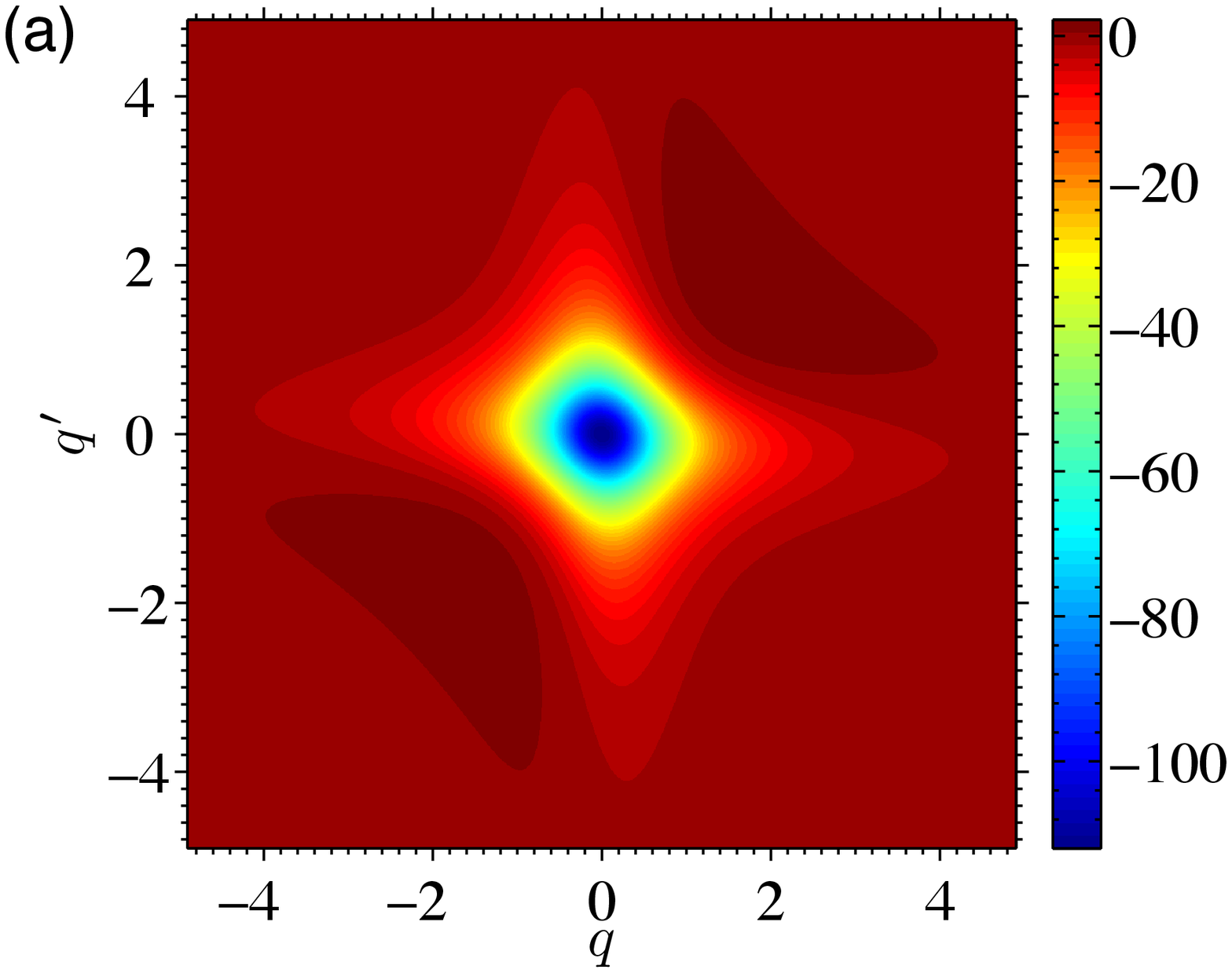}\\
~~\includegraphics[width=7.25cm]{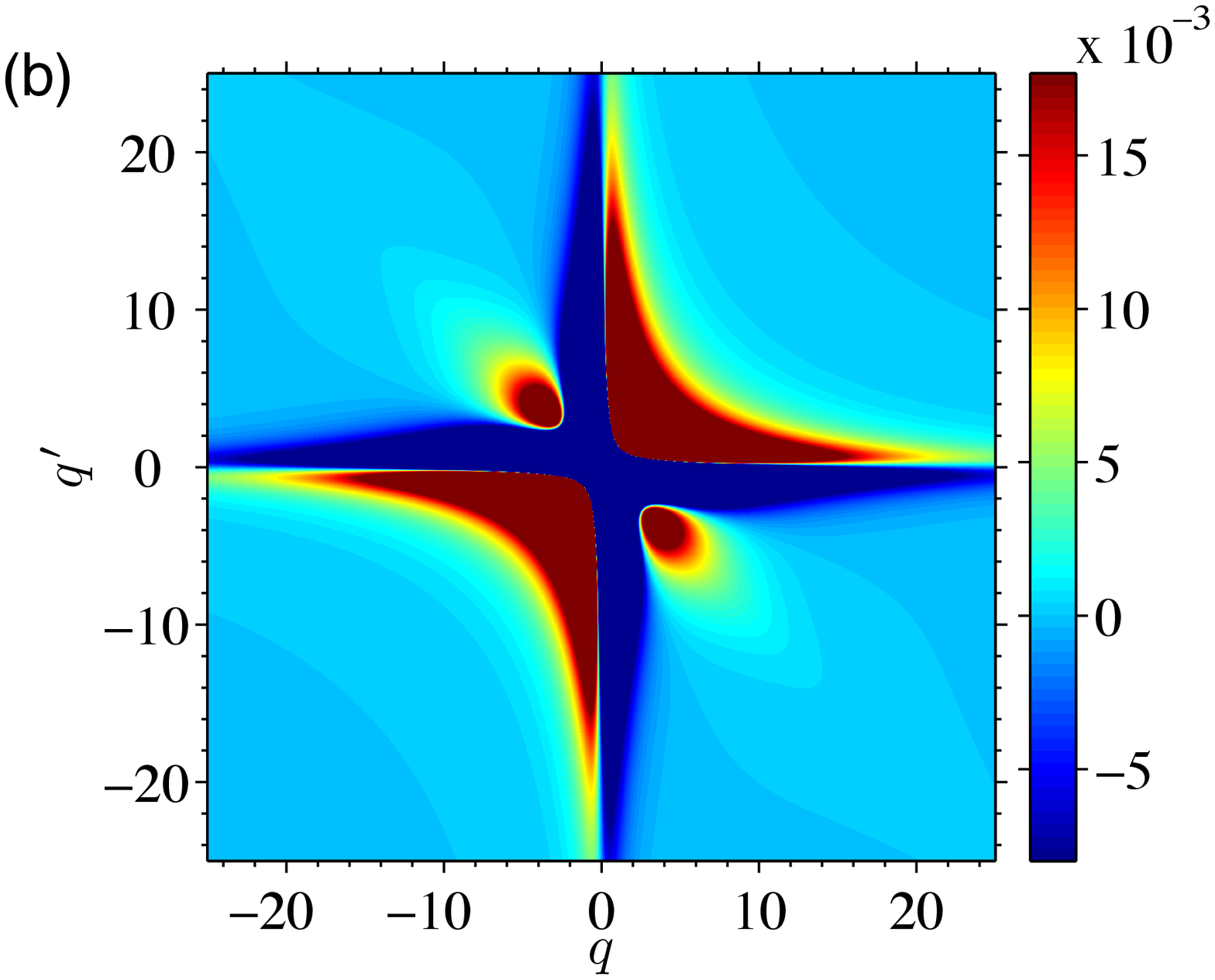}\\
{\hspace{-0.5cm}\includegraphics[width=7.8cm]{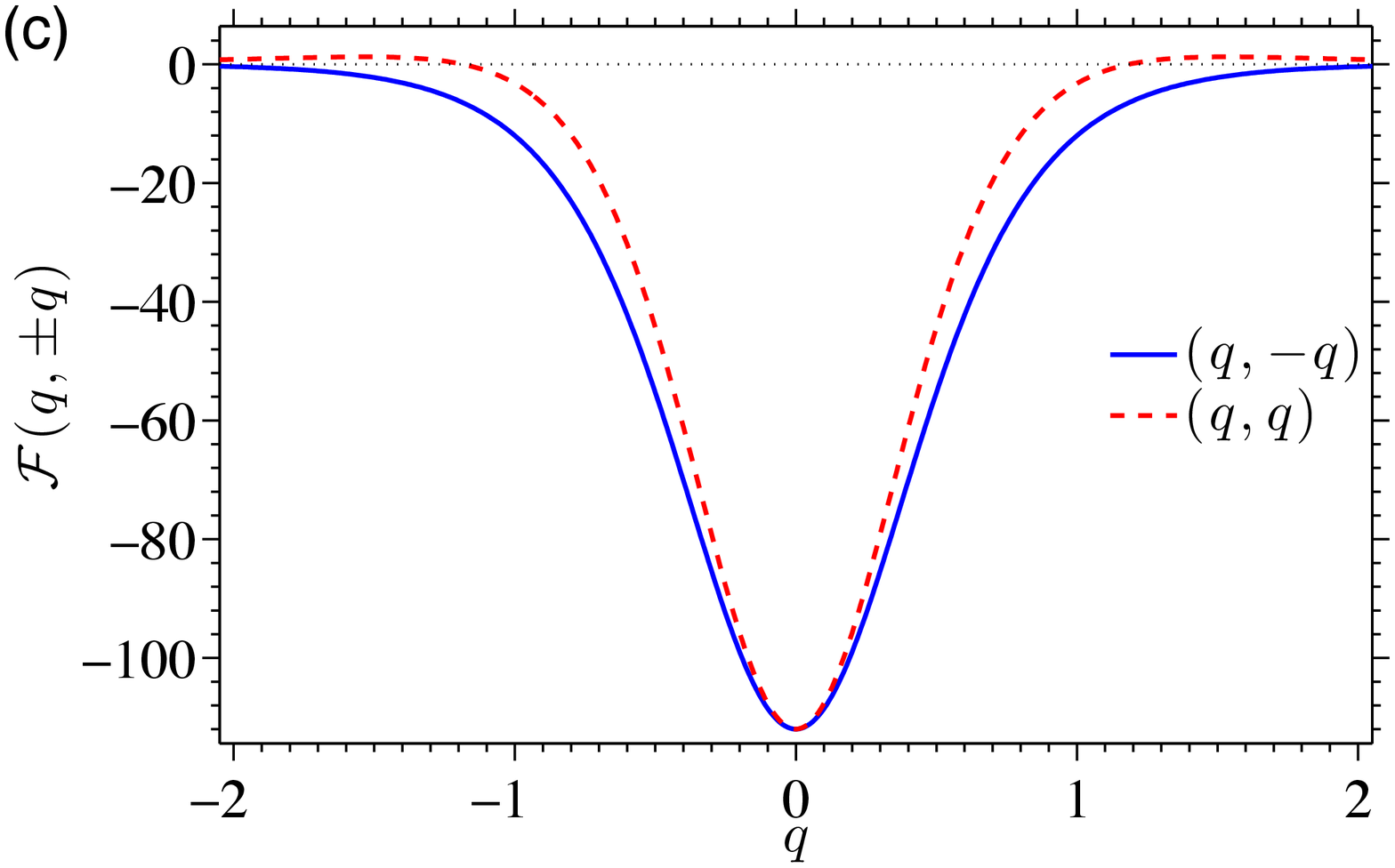}}
\caption{(Color online) (a) Dimensionless regular part of the (unnormalized) two-body momentum correlation function,
${\cal F}
(q,q^{\prime})$, given by Eq.~(\ref{eq:fqbec}), of a uniform 1D Bose gas in the quasicondensate regime. (b) Same as in (a), but showing the details at small correlation amplitudes (see the scale on the color bar) and in a larger window of values of $(q,q^{\prime})$. The small negative and positive amplitudes seen here get ``magnified'' when the function ${\cal F}(q,q')$ is normalized to $\langle \hat{n}_k\rangle \langle \hat{n}_k'\rangle$, as is done in Fig. \ref{fig.dnquasibec-norm}. (c) Function ${\cal F}(q,q^{\prime})$ along the diagonal ($q=q^{\prime}$) and antidiagonal ($q=-q^{\prime}$).
}
\label{fig.dnquasibec}
\end{figure}

Th evaluation of the integral on the left-hand side of Eq.~(\ref{eq:sumrule1})
gives the value of $-8$, implying that the sum rule is
indeed approximately satisfied as long as $T\ll T_{\rm{co}}$, i.e., deep in
the quasicondensate regime.  On the other hand, as the temperature increases
and approaches the quasicondensation crossover $T_{\rm{co}}$, the term
$(T/T_{\rm{co}})^2$ on the right-hand side of Eq.~(\ref{eq:sumrule1})
  becomes non-negligible, implying that our result for the pair-correlation
function $\widetilde{\cal G}(k,k')$, given by Eqs. (\ref{eq.Fronde}) and
(\ref{eq:fqbec}), is no longer valid as it fails to satisfy the sum rule
\cite{sum-rule-consistency}. The physical origin of this failure lies in the
fact that the density fluctuations at temperatures near $T_{\rm{co}}$ are no
longer negligible.

The two-body correlation function ${\cal G}(k,k^{\prime})$, given by
Eq.~(\ref{eq.GvsGtilde}), in the quasicondensate regime, of which the regular
part $\widetilde{{\cal G}}(k,k^{\prime})$ is described by the universal
dimensionless function $\mathcal{F}(q,q^{\prime})$, is one of the key results
of this paper.  The function $\mathcal{F}(q,q^{\prime})$ is shown in
Figs.~\ref{fig.dnquasibec}(a)--(c); as $\mathcal{F}(q,q^{\prime})$ is
  independent of the system size $L$, it essentially describes the
(unnormalized) two-body momentum correlations in the thermodynamic limit.  As
we see, the correlation function is nonzero on the entire 2D plane of momentum
pairs $(k,k^{\prime})$; this can be contrasted with the singular behaviorof
correlations in the true condensate where ${\cal G}(k,k^{\prime})$ was nonzero
only for $k^{\prime}=\pm k$. This effective broadening of correlations is the
first consequence of large phase fluctuations in the quasicondensate
regime compared to the behaviorin the true condensate. Next, $\widetilde{{\cal G}}(k,k^{\prime})$ and ${\cal
  P}(k)$ both scale as $l_{\phi}/L$ and therefore are vanishingly small as
$L\gg l_{\phi}$. For $k=-k^{\prime}$, this means that the perfect
opposite-momentum correlations [${\cal P}(k)=1$], which were present in
  the true condensate, essentially disappear in the quasicondensate regime.
Finally, we find \textit{negative} correlations (or anticorrelations)
in $\widetilde{{\cal G}}(k,k^{\prime})$; these are pronounced mostly in the
regions of $k^{\prime}k<0$ [see Fig.~\ref{fig.dnquasibec}(b)]. The
correlations fall to zero on a typical scale of $k\sim 1/l_{\phi}$
($q=2l_{\phi}k\sim 1$). This is expected, as $l_{\phi}$ is the length scale
governing the first-order spatial correlation function $G_1(z_1,z_2)$ in the
quasicondensate regime, and the momentum correlations depend only on
$G_1(z_1,z_2)$ in this regime.

\begin{figure}[tbp]
\includegraphics[width=6.45cm]{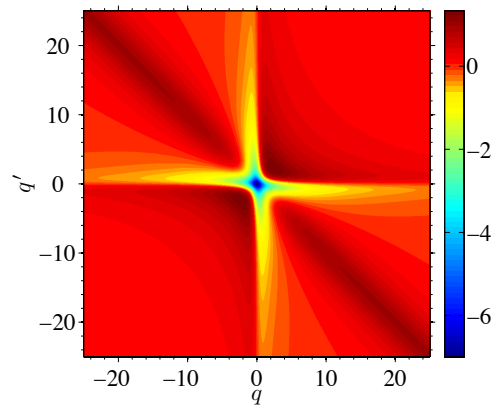}
\caption{(Color online) Normalized regular part of the two-body correlation function
$f(q,q')$ as a function of the dimensionless momenta $q=2l_{\phi}k$ and $q^{\prime}=2l_{\phi}k'$.
}
\label{fig.dnquasibec-norm}
\end{figure}

To gain further insights into the strength of the two-body correlations, we
consider the \textit{normalized} regular part of the two-body correlation
function,
\begin{equation}
\widetilde{g}^{(2)}(k,k^{\prime})\equiv\frac{\widetilde{{\cal G}}(k,k')}{\langle\hat{n}_{k}\rangle\langle\hat{n}_{k^{\prime}}\rangle}.
\end{equation}
Using Eqs.~(\ref{eq.Fronde}) and (\ref{eq.Lorentzian}), this can be
rewritten as
\begin{equation}
\widetilde{g}^{(2)}(k,k^{\prime})=\frac{l_{\phi}}{L}\,f(2 l_{\phi}k,2l_{\phi}k'),
\label{eq.g2qBEC}
\end{equation}
where 
\begin{equation}
f(q,q')=\frac{\mathcal{F}(q,q')}{16}\left(1+q^{2}\right)\left(1+q^{\prime 2}\right).
\label{eq.f}
\end{equation}
is a dimensionless universal function describing the two-body correlations of
a 1D quasicondensate in the thermodynamic limit. The function $f(q,q')$ is
plotted in Fig~\ref{fig.dnquasibec-norm}.  As we see, the normalization leads,
at $k\gg 1/l_{\phi}$, to the recovery [cf. Fig. \ref{fig.dnquasibec}(b)] of
positive correlations around the the antidiagonal $k^{\prime}=-k$, predicted
by the simple model of Sec. \ref{sec:simple}. These correlations can be also
thought of as the remnants of the nearly perfect correlations in a true
condensate at $k\ll 1/\xi$.

Finally, we note that $\widetilde{g}^{(2)}(k,k^{\prime})$ can be related
  to Glauber's normally-ordered second-order correlation function
\begin{equation}
g^{(2)}(k,k^{\prime})=\frac{\langle \hat{\psi}_{k}^{\dagger}\hat{\psi}_{k^{\prime}}^{\dagger}\hat{\psi}_{k^{\prime}}\hat{\psi}_{k}\rangle }{\langle \hat{\psi}_{k}^{\dagger}\hat{\psi}_{k}\rangle \langle \hat{\psi}_{k^{\prime}}^{\dagger}\hat{\psi}_{k^{\prime}}\rangle }.
\end{equation}
Indeed, by reordering the creation and annihilation operators, we can first
express the $g^{(2)}(k,k^{\prime})$ function in terms of the correlation
function ${\cal G}(k,k')$, given by Eq.~(\ref{eq.Gkkp}):
\begin{equation}g^{(2)}(k,k^{\prime})=1-\frac{1}{\langle\hat{n}_{k}\rangle}\delta_{k,k^{\prime}}+\frac{{\cal G}(k,k')}{\langle\hat{n}_{k}\rangle\langle\hat{n}_{k^{\prime}}\rangle}.
\end{equation}
Using now the general structure of ${\cal G}(k,k')$ from Eq.~(\ref{eq.GvsGtilde}) (valid for $L\gg l_{\phi}$) we obtain
\begin{equation}
g^{(2)}(k,k^{\prime})=1+\delta_{k,k^{\prime}}+\widetilde{g}^{(2)}(k,k^{\prime}).
\end{equation}
Here, the first term corresponds to uncorrelated atoms, the second term is the
bosonic bunching term, and the last term is the normalized regular part
corresponding to $\widetilde{{\cal G}}(k,k')$ given by Eq.~(\ref{eq.g2qBEC}).

According to our results, the normally ordered normalized correlation function for equal momenta is given by
\begin{equation}
g^{(2)}(k,k)=2+\widetilde{g}^{(2)}(k,k)=2+{\cal O}(l_{\phi}/L),
\label{eq:g2bunching}
\end{equation}
while for opposite momenta it is given by
\begin{equation}
g^{(2)}(k,-k)=1+\widetilde{g}^{(2)}(k,-k)=1+{\cal O}(l_{\phi}/L).
\label{eq:g2uncorrelated}
\end{equation}
The small contributions ${\cal O}(l_{\phi}/L)$ are described by
Eq. (\ref{eq.g2qBEC}) and are, in principle, detectable using the precision of
currently available experimental techniques. Apart from the need for high precision on the signal, 
resolving 
the shape of the $\widetilde{g}^{(2)}(k,k')$-function requires experimental 
momentum resolution better than the separation $\Delta k=2\pi/L$ between the individual
momentum states (for resolutions that are insufficient to resolve separations of
$\sim1/L$, see Sec. \ref{sec:experimental}).

As we see from Eq.~(\ref{eq:g2bunching}), the amplitude of equal-momentum
correlations is close to the pure thermal bunching level of
$g^{(2)}(k,k)=2$, implying large momentum-space density fluctuations. The
nearly thermal level of correlations here is due to the large phase
fluctuations present in a 1D quasicondensate. This makes the equal-momentum
correlations analogous to those of a `speckle' pattern \cite{Truscott:speckle} where
many sources with random phases contribute to the familiar Hanbury
Brown--Twiss interference \cite{HBT}. We emphasize, however, that the nearly
thermal equal-momentum correlations are obtained here for a quasicondensate,
which should be contrasted to the uncorrelated level of the two-point correlation
function in position space \cite{KheruntsyanPRL03,Sykes:08},
$g^{(2)}(z,z)\simeq 1$, due to the \textit{suppressed} real-space density
fluctuations. Equation (\ref{eq:g2uncorrelated}), on the other hand, shows
that the opposite-momentum correlations are close to the uncorrelated level of
$g^{(2)}(k,-k)\simeq 1$, which is in contrast to the strong respective
correlations [$g^{(2)}(k,-k)= 2+1/\langle \hat{n}_k \rangle$ at $T=0$, 
and $g^{(2)}(k,-k)\simeq 2$ at finite $T$ for $\langle \hat{n}_k \rangle \gg 1$]
present in a true condensate. As we mentioned earlier, the
opposite-momentum correlations are essentially destroyed by the phase
fluctuations. Finally, a significant region of pairs of momenta $k'\neq k$
around the origin shows a small degree of anticorrelation, $g^{(2)}(k,k')< 1$,
which was not \textit{a priori} expected.

\section{From the quasicondensate to the ideal Bose gas regime}
\label{sec:crossover}

\subsection{Classical field approach}
\label{sec:classfield}

In the quasicondensate regime, $T\ll T_{\rm{co}}$, higher-order correlation
functions can always be expressed in terms of the first-order correlation function as
in Eq.~(\ref{eq.g2quasicond}). Therefore, ${\cal G}(k,k')$ in
Eq. (\ref{eq.Gkkpcorr}) contains the same information as the momentum
distribution $\langle \hat{n}_k\rangle$, given by Eq. (\ref{eq:nk}).  In particular,
the dependence on the temperature comes about only through the phase
correlation length $l_{\phi}$. This is, however, no longer true when the
temperature becomes of the order of the crossover temperature $T_{\rm{co}}$,
in which case the physics depends not only on the phase fluctuations, but
also on the density fluctuations.

To compute the correlation functions at $T\gtrsim T_{\rm{co}}$, we resort to
the classical field (or $c$-field) approach of Ref.
\cite{Castinatomlaser}. In this approach, the quantum field operators
$\hat{\psi}$ and $\hat{\psi}^\dagger$ are approximated by $c$-number fields
$\psi$ and $\psi^*$, whose grand-canonical partition function (in a
path-integral formulation) is given by
\begin{equation}
{\cal Z}=\!\!\int \!\!{\cal D}\psi {\cal D}\psi^{*}\exp\left(-\frac{1}{k_BT}\int_{0}^{L}\!\!dz\: 
\mathcal{H}_c\right).
\label{eq.Zclass}
\end{equation}
Here the function $\mathcal{H}_c\left(\psi,\psi^{*}\right)$ is obtained from the
Hamiltonian density~(\ref{eq.HLL}) by replacing the operators with
$c$-fields. The classical field approach is expected to be valid
for high occupancy of the low-momentum modes contributing to the momentum
correlation function. This condition is satisfied in a broad range of
temperatures, including in the quasicondensate regime,
$g\rho e^{-2\pi/\sqrt{\gamma}}<k_BT<\sqrt{\gamma}\hbar^2\rho^2/m$ \cite{quantum_fluctuations,regimes}, and up
to the temperatures corresponding to the degenerate ideal Bose gas regime,
$\sqrt{\gamma}\hbar^2\rho^2/m< k_BT < \hbar^2 \rho^2/m$ 
\cite{Castinatomlaser}.

It is convenient to introduce a dimensionless field $\widetilde \psi=\psi/\psi_0$ and a dimensionless coordinate 
$s=z/z_0$, with 
 \begin{equation}
   \label{eq:rescaling}
 \psi_0=\left(\frac{m k_B^2T^2}{\hbar^2 g}\right)^{1/6}, \quad  z_0= \left(\frac{\hbar^4}{m^2 g k_BT}\right)^{1/3},
  \end{equation}
and rewrite the effective `action' in Eq.~(\ref{eq.Zclass}) in the dimensionless form:
\begin{align}
\label{eq:weight}
\frac{1}{k_BT} \int_{0}^{L}  \!dz\mathcal{H}_c
=\!\int_{0}^{L/z_0} \! ds \left( \frac{1}{2} |\partial _s\widetilde\psi|^2
+ \frac{1}{2} |\widetilde\psi|^4  
  -{\eta} |\widetilde\psi|^2\right).
\end{align}
This form of the action is controlled by a
single dimensionless parameter
\begin{eqnarray}
\label{eq:eta}
\eta = \left ( \frac{\hbar^{2} }{mg^2 k_B^2T^2}\right )^{1/3}\mu.
\end{eqnarray}

Because of the scaling  relations (\ref{eq:rescaling}), 
the density $\rho=\langle \psi^{*}\psi\rangle$ can be written as $\rho=h(\eta)(mk_B^2T^2/\hbar^2g)^{1/3}$ using
a dimensionless function $h(\eta)\equiv \langle \widetilde\psi^{*}\widetilde\psi\rangle$.
 Similarly, the phase correlation length $l_\phi$, given by Eq.~(\ref{eq.lphi}), can be written as  $l_\phi=z_0h(\eta)$. 
Thus, the length scale $z_0$ can be replaced by  $l_\phi$ and 
therefore the one- and two-body correlation functions in 
Eqs.~(\ref{eq:g1}) and (\ref{eq:g2}) scale as
\begin{eqnarray}
  \label{eq:gscale}
  G_1(z_1,z_2) &=& \rho\; h_1\left(\frac{z_1}{l_\phi},
    \frac{z_2}{l_\phi};\eta\right), \\
  G_2(z_1,z_2,z_3,z_4) &=& \rho^2\; h_2
  \left(\frac{z_1}{l_\phi},\frac{z_2}{l_\phi},
    \frac{z_3}{l_\phi},\frac{z_4}{l_\phi};\eta\right),
\end{eqnarray}
where  $h_1$ and $h_2$ are  dimensionless functions.

By substituting these scaled correlation functions into Eq.~(\ref{eq.Gkkpcorr}) and using the
fact that the integrand is invariant by a global translation of the
coordinates, we find 
\begin{equation}
\widetilde{\cal G}(k,k') = \frac{l_\phi}{L}
(\rho l_\phi)^2 {\cal F}(2l_{\phi} k,2l_\phi k';\eta),
\label{eq.Frondeeta}
\end{equation}
where ${\cal F}(q,q';\eta)$ is a dimensionless function parametrized by $\eta$. This relation 
generalizes
Eq.~(\ref{eq.Fronde}) beyond the quasicondensate regime, with the departure being characterized by the value of 
$\eta$ (see below).

To find ${\cal F}(q,q';\eta)$, we still need to calculate the dimensionless function $h$ and correlations $h_1$ and $h_2$
for the rescaled fields $\widetilde\psi$ and $\widetilde\psi^*$, with the action
given by
Eq.~(\ref{eq:weight}). 
As shown in Ref.~\cite{Castinatomlaser}, this $c$-field problem
can be mapped into the quantum-mechanical problem of a particle moving in an external potential.  
More precisely, expressing the action
(\ref{eq:weight}) in terms of the
real and imaginary components of $\widetilde\psi =x+iy$ and interpreting $s$ as the 
imaginary time, the problem can be mapped to the quantum mechanics of a particle in two dimensions with the Hamiltonian
\begin{eqnarray}
{H} \!=\! \frac{1}{2}\left(p_x^2+p_y^2\right)+ 
\frac{1}{2}\left(x^2+y^2\right)^2\!
-\eta\left(x^2+y^2\right).
\label{eq:ham_qm}
\end{eqnarray}
Calculating the eigenvalues and matrix elements of this Hamiltonian allows one 
to compute the correlation function ${\cal F}(q,q';\eta)$. This is done in Appendix \ref{sec:cfham}.

\subsection{Correlations in the crossover region}
\label{sec:IBG}

The power of the $c$-field approach lies in the ability to describe the momentum correlations not 
only in the quasicondensate regime, but also in the entire crossover region 
between the quasicondensate and the degenerate ideal Bose gas. As shown in Appendix \ref{sec:cfclquasi}, in the quasicondensate regime where $\eta \gg 1$ (corresponding to a positive chemical potential $\mu$), we recover the results of Sec.~\ref{sec:Luttinger}, with Eq.~(\ref{eq:fqbec}) referring to $\mathcal{F}(q,q';+\infty)\equiv \mathcal{F}(q,q')$. The opposite limit $\eta\ll-1$ corresponds 
to the degenerate ideal Bose-gas
regime with negative $\mu$. In this case, the quartic term in the Hamiltonian (\ref{eq:ham_qm}) has a negligible effect
on the lowest energy eigenstates and the problem is reduced to a simple two-dimensional harmonic
oscillator. As shown in Appendix~\ref{sec:idealgas}, in this limit, 
we obtain the ideal Bose-gas result of ${\cal F}(q,q';-\infty)=0$.

\begin{figure}
\includegraphics[width=7.4cm]{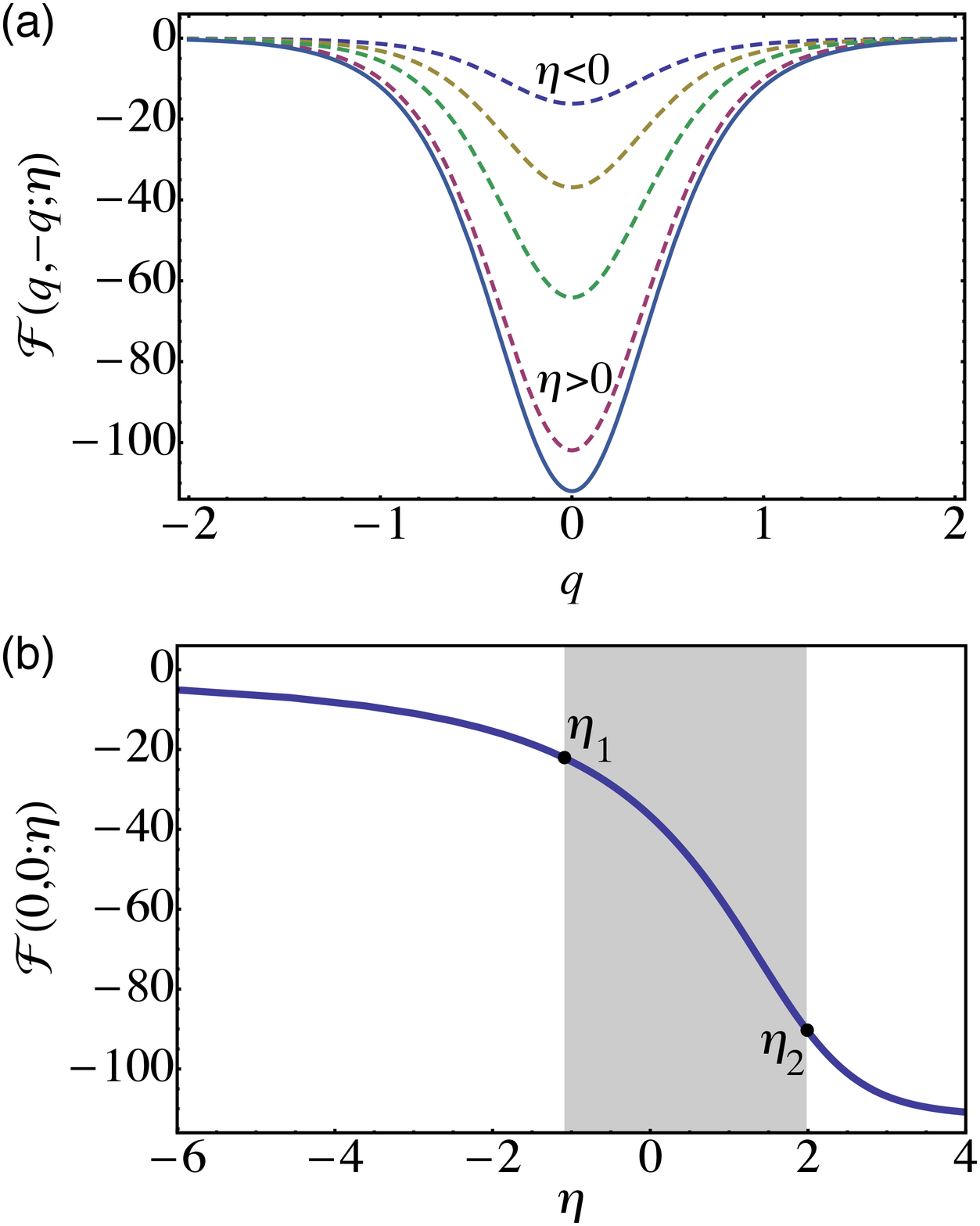}
\caption{(Color online) (a) The antidiagonal correlation function 
$F(q,-q;\eta)$, for (from top to bottom, dashed lines) $\eta=-1.87, 0, 1.12$, and $2.56$. The lowest (solid)
curve is for the limiting quasicondensate regime, $\mathcal{F} 
(q,-q)\equiv \mathcal{F}(q,-q; +\infty)$, described by Eq.~(\ref{eq:fqbec}) and shown in 
Fig.~\ref{fig.dnquasibec}~(c).
(b) The minimum value of $F(q,-q;\eta)$ as a function of $\eta$. The shaded area shows the crossover region between 
$\eta_1 < \eta < \eta_2 $ (see text).
}
\label{fig.figfandD}
\end{figure}

In Fig.~\ref{fig.figfandD}(a) we show how the antidiagonal correlation function $\mathcal{F}(q,-q;\eta)$ 
changes from its quasicondensate value of Eq.~(\ref{eq:fqbec}) to zero as
$\eta$ is continuously changed from $+\infty$ to $-\infty$. 
To quantify the width of the crossover in terms of $\eta$,
we consider the peak value of the correlation function, $\mathcal{F}(0,0;\eta)$, 
and plot it as a function of $\eta$ in Fig.~\ref{fig.figfandD}(b). As we see,
$\mathcal{F}(0,0;\eta)$ goes from its minimum value of about $-112$ in the quasicondensate 
regime ($\eta\gg 1$) to zero in the ideal Bose gas regime ($\eta \ll -1$).

We can define the crossover region to correspond to 
$\eta_1<\eta<\eta_2$, where the bounds $\eta_1$ and $\eta_2$ are chosen, respectively, at $20\%$ and $80\%$ of the value of $\mathcal{F}(0,0;\eta)$ in the quasicondensate regime; our numerical solutions give then 
$\eta_1\simeq -1.1$ and $\eta_2\simeq 2.0$. Recalling that the dimensionless parameter $\eta$ is defined via Eq. (\ref{eq:eta}), this can be converted into the crossover bounds on the chemical potential, $\eta_1< \mu/\mu_{\rm{co}}< \eta_2$, where $\mu_{\rm{co}}\equiv k_BT(mg^2/\hbar^2k_BT)^{1/3}$ \cite{Bouchoule07}. Similarly, recalling that the density $\rho$ was determined by the dimensionless function $h(\eta)$, via $\rho=h(\eta)(mk_B^2T^2/\hbar^2g)^{1/3}$, we can use the numerically found values of $h(\eta)$ to rewrite the crossover bounds in terms of the density as 
$0.5 < \rho/\rho_{\rm{co}} < 1.6$, where $\rho_{\rm{co}}\equiv (mk_B^2T^2/\hbar^2g)^{1/3}$ \cite{Bouchoule07}.

\section{Experimental considerations}
\label{sec:experimental}

The results obtained so far are directly applicable to experimentally measured momentum distributions and correlation functions as long the momentum resolution is sufficient to resolve the individual momentum states separated by $\Delta k=2\pi/L$. However, typical resolution in ultracold-atom experiments is insufficient to resolve momentum scales of the order of $1/L$. Because of this, the measured signal corresponds to integrated atom-number counts $N_k$ in individual detection ``bins'' (such as camera pixels in absorption imaging) corresponding to the momentum $k$. 

To address the situation with low momentum resolution and relate our calculated correlation functions to the experimentally accessible quantities, we assume that the detection bin size $\Delta_{k}$ in momentum space fulfills
$\Delta_{k} \gg 1/L$. In addition, we assume that $\Delta_{k} \ll 1/l_{\phi}$, so that the bulk of the 
momentum distribution is still well resolved. With these assumptions, the average (over many experimental runs) atom number in a bin $\langle N_k \rangle$ is related to the original average mode occupation number $\langle \hat{n}_k\rangle$ via $\langle N_k \rangle=(L\Delta_k/2\pi)\langle \hat{n}_k\rangle$, i.e., it accounts for a factor equal to the number of original momentum states contributing to the bin, $\Delta_k/\Delta k=L\Delta_k/2\pi$.
Next, the average correlation between the bin population fluctuations is related to the correlation function ${\cal G}(k,k')$, given by Eqs.~(\ref{eq.GvsGtilde}) and (\ref{eq.Fronde}), via 
\begin{align}
\label{eq.correlpixels}
\langle N_{k}N_{k'}\rangle-\langle N_{k}\rangle\langle N_{k'}\rangle =\langle N_k\rangle\delta_{k,k'} \qquad\qquad\qquad\quad \nonumber \\
\qquad + \langle N_{k}\rangle\langle N_{k'}\rangle \left[\frac{2\pi}{L\Delta_{k}} \delta_{k,k'} 
+ \frac{l_{\phi}}{L}f(2l_\phi k,2l_\phi k')\right],
\end{align}
where the universal function $f(q,q')$ is given by Eq. (\ref{eq.f}). 
In Eq. (\ref{eq.correlpixels}), the 
first term is the shot noise, the second term corresponds to the bunching 
term in Eq. (\ref{eq.GvsGtilde}),  
and the third term is the contribution of the regular part, $\widetilde{{\cal G}}(k,k')$.

The shot-noise term 
is much smaller than the 
bunching term as long as highly populated momentum states 
are considered, i.e., $\langle \hat{n}_k\rangle\gg 1$. 
The latter condition is satisfied for momenta $k\lesssim 1/l_\phi$ 
(containing the bulk of the momentum distribution), and therefore the shot-noise 
term can be safely neglected for these momenta.

Comparing now the bunching term and the regular component 
(with the comparison being relevant only for $k=k'$), we see that they both scale 
inversely proportionally to the system size $L$ (and therefore are small), but the 
regular component is much smaller than the bunching term as the dimensionless function $f(2l_\phi k,2l_\phi k')$
is of the order of one and we have assumed $\Delta_{k} \ll 1/l_{\phi}$. However, the ratio of these 
two terms is independent of $L$ and therefore is finite in the thermodynamic limit. 
As this ratio is proportional to $\Delta_{k} l_{\phi}\ll 1$, detecting the contribution 
of the regular component is going to depend on actual experimental parameters and 
the precision (signal-to-noise) with which the atom-number fluctuations can be measured. 
High-precision measurements of atom-number fluctuations, capable of resolving 
small signals like this or even below the shot-noise level, have been demonstrated in many ultracold atom experiments \cite{Sanner10,Mueller10,ArmijoSkew,Armijo10_2,Subbunching,Sanner:11,ScaleInvariance2,Manz2010}.

Considering now the cross correlation between atom-number counts in different bins, $k'\neq k$, we see that the only contribution to Eq. (\ref{eq.correlpixels}) comes from the regular component. This scales as $1/L$, but again such a magnitude of the cross correlation should be accessible with state-of-the-art measurement techniques, as demonstrated, e.g., in Ref. \cite{ArmijoSkew}.

\section{Summary}
\label{sec:summary}

To summarize, we have calculated the two-body momentum correlations for a weakly interacting, uniform 1D Bose gas. Our results span the entire quasicondensate regime, where the correlations are derived analytically in terms of a universal dimensionless function $f(2l_\phi k,2l_\phi k')$, as well as the crossover to the ideal Bose-gas regime, where the correlations are calculated numerically using the classical field method. A natural extension of the approaches employed here would be the calculation of these correlations for harmonically trapped gases \cite{IBG},
which is more appropriate for quantitative comparisons with experiments beyond the widely used local-density approximation. Calculating and understanding the momentum correlations in the strongly interacting regimes would require the development of alternative theoretical approaches and remains an open problem. The knowledge of such correlations is important in the studies of nonequilibrium dynamics from a known initial state and the subsequent thermalization in isolated quantum systems \cite{Dima:08,Muth:10,Giraud:thermalization,Polkovnikov:review}.

\begin{acknowledgments}
The authors thank T. M. Wright for critical reading of the manuscript and G. V. Shlyapnikov for stimulating discussions in the early stages of this work.
I.B. acknowledges support from the Triangle de la Physique, the ANR
Grant No. ANR-09-NANO-039-04, and the 
Austro-French FWR-ANR Project No. I607.
K.V.K. acknowledges support by the ARC Discovery Project
Grant No. DP110101047. M.A. and D.M.G. acknowledge support by the EPSRC, and the hospitality of the Institut d'Optique during their visit.

\end{acknowledgments}

\appendix

\section{Classical field approach: diagonalizing the effective hamiltonian}
\label{sec:cfham}

In this appendix, we outline how the classical field correlation functions can be
computed using the the equivalent quantum-mechanical problem of a 
particle in an external potential. We recall that the classical-to-quantum mapping 
is done by expressing the $c$-field $\psi=x+iy$ via its real and imaginary parts
which, in turn, are treated as coordinates of a quantum-mechanical particle in
imaginary time with the Hamiltonian given by Eq.~(\ref{eq:ham_qm}). Here we imply
the scaling of Eq.~(\ref{eq:rescaling}) but omit the tilde on top of the
coordinates and $c$-fields for the sake of notational simplicity.

Using the notations of effective quantum mechanics, the first- and 
second-order correlation functions of the $c$-fields, 
\begin{equation}
G_1(s_1,s_2)=\left<\psi^{*}\left(s_{1}\right)\psi\left(s_{2}\right)\right>
\end{equation}
and
\begin{equation}
G_2(s_1,s_2,s_3,s_4)=\left<\psi^{*}\left(s_{1}\right)\psi\left(s_{2}\right)\psi^{*}\left(s_{3}\right)\psi\left(s_{4}\right)\right>,
\end{equation}
are given by
\begin{equation}
G_1=\frac{Tr[U_{L-s'_{1}}\Psi_1 U_{s'_{1}-s'_{2}}
\Psi_2 U_{s'_{2}}]}{Tr[U_{L}]}
\label{eq.c2CF}
\end{equation}
and
\begin{equation}
G_2=\frac{Tr[U_{L-s'_{1}}\Psi_1U_{s'_{1}-s'_{2}}
\Psi_2 U_{s'_{2}-s'_{3}}\Psi_3U_{s'_{3}-s'_{4}}\Psi_4 
U_{s'_{4}}]}{Tr[U_{L}]},
\label{eq.c4CF}
\end{equation}
where we have omitted the arguments of $G_1$ and $G_2$ for notational brevity.
Here $U_{s}=e^{-s H}$ is the imaginary-time evolution operator generated
by the Hamiltonian (\ref{eq:ham_qm}). In Eq.~(\ref{eq.c4CF}), 
we take into account the automatic time ordering implied by the path
integral by introducing $s'_1\geq...\geq s'_4$, which is the ordered permutation of
$s_1,...,s_4$. The operator {$\Psi_k$} stands for $\psi=x+iy$ if $s'_k$ equals
$s_2$ or $s_4$, or for $\psi^*=x-iy$ if $s'_k$ equals $s_1$ or $s_3$.

In the limit $L\rightarrow \infty$, the ground state $|0\rangle$ 
gives the dominant 
contribution to both the numerator and denominator in Eq. (\ref{eq.c4CF}), 
and hence the correlation functions 
reduce to 
\begin{equation}
G_1=\left\langle 0 | \Psi_1U_{s'_{1}-s'_{2}}\Psi_2  |0\right \rangle
\label{eq.g1class}
\end{equation}
and
\begin{equation}
G_2=\left\langle 0 | \Psi_1U_{s'_{1}-s'_{2}}
\Psi_2 U_{s'_{2}-s'_{3}}\Psi_3
U_{s'_{3}-s'_{4}}\Psi_4  |0\right \rangle,
\label{eq.g2class}
\end{equation}
where we have set the ground state energy to $\epsilon_0=0$.

The expectation values on the right-hand sides of Eqs.~(\ref{eq.g1class}) and (\ref{eq.g2class}) are best evaluated
in the eigenbasis of the Hamiltonian $H$. Let $|\alpha\rangle$ be the set of
eigenstates of $H$ 
with energy eigenvalues $\epsilon_\alpha$,
\begin{equation}
{ H}|\alpha\rangle=\epsilon_\alpha|\alpha\rangle.
\end{equation}
The eigenstates $|\alpha\rangle = |n,m\rangle$ are classified by the principal
($n$) and angular momentum ($m$) quantum numbers such that  
in polar coordinates, the
eigenfunctions 
\begin{equation}
\langle r,\theta|\alpha\rangle=\frac{1}{\sqrt{2\pi}}\phi_{n} ^{m}(r)e^{im\theta}
\end{equation}
obey the following eigenvalue equation:
\begin{equation}
\left[-\frac{1}{2r}\frac{\partial}{\partial r}\left(r\frac{\partial}{\partial
      r}\right)+\frac{m^2}{2r^2}+\frac{r^{4}}{2}-\eta
  r^{2}\right]\phi_n^m(r)=\epsilon_n^m\phi_n^m(r)\, .
\label{eq.polarSchroedinger}
\end{equation}

In terms of the matrix elements 
\begin{eqnarray}
  \label{eq:matrixel}
A_{\alpha\beta}=\langle\alpha|\psi|\beta\rangle=\langle\alpha|x+iy|\beta\rangle, 
\end{eqnarray}
the correlation functions are 
\begin{equation}
G_1(s_1,s_2)=\sum_{\alpha} e^{-|s_1-s_2|(\epsilon_\alpha-\epsilon_0)} |A_{\alpha0}|^2
\label{eq.g1classexpand}
\end{equation}
and 
\begin{equation}
G_2(s_1,s_2,s_3,s_4)=\sum_{\alpha\beta\gamma} 
e^{-K} A_{\alpha0}^*A_{\alpha\beta}
A_{\gamma\beta}^*A_{\gamma0},
\label{eq.g2classexpand}
\end{equation}
for the case $s_1>s_2>s_3>s_4$, where we have defined
\begin{eqnarray}
K&=&\epsilon_0(s_4-s_1) +\epsilon_\gamma(s_3-s_4)\nonumber \\
 &+& \epsilon_\beta(s_2-s_3)+\epsilon_\alpha(s_1-s_2).
\end{eqnarray}
For different orderings of $s_1, s_2, s_3$, and $s_4$, similar expressions can be
obtained. Although there are, in general, $4!=24$ cases to consider, by noticing
that the expectation value for $G_2(s_1,s_2,s_3,s_4)$ remains invariant under the exchange 
of $s_{1}\!\rightleftarrows \!s_3$, or $s_{2}\!\rightleftarrows \!s_4$, 
and also under the simultaneous exchange of $s_{1}\!\rightleftarrows \!s_2$ 
and $s_{3}\!\rightleftarrows \!s_4$, we realize that it is sufficient 
to compute $G_2(s_1,s_2,s_3,s_4)$
in just three cases: $s_{1}>s_{2}>s_{3}>s_{4}$, $ s_{1}>s_{2}>s_{4}>s_{3} $,
and $ s_{1}>s_{3}>s_{2}>s_{4}.$ The remaining 21 expressions can be obtained
from these using symmetry considerations.

Solving the Schroedinger equation (\ref{eq.polarSchroedinger}) numerically and
evaluating the matrix elements given by Eq.~(\ref{eq:matrixel}) yields the correlation
functions in Eqs. (\ref{eq.g1class}) and (\ref{eq.g2class}). It should be noted
that the sums in Eqs.~(\ref{eq.g1classexpand}) and (\ref{eq.g2classexpand})
contain only a finite number of terms because of the selection rule,
$A_{\alpha\beta}\propto\delta_{m_\alpha, m_\beta+1}$, and the fact that the bra-ket states
with a very large difference in the respective values of $n$ give negligible matrix 
elements due to very different nodal
structure.

\section{Quasicondensate limit}

\label{sec:cfclquasi}

In this appendix, we show that the classical field approximation correctly
predicts the correlation functions in Eqs.~(\ref{eq.g1quasicond})
and~(\ref{eq.Cft}), in the limit $\eta\gg 1$.  
In this limit, the wavefunctions
of the lowest-lying states  differ from zero significantly  only for 
$r \simeq r_0=\sqrt{\eta}$, so that the Hamiltonian becomes separable 
into the azimuthal and radial degrees of freedom. This has two consequences
on the classical field calculations. 

First,
the wave functions $\phi_n^m(r)$ are approximately
independent of $m$, while the azimuthal kinetic energy is reduced 
to
\begin{equation}
{H} \simeq-\frac{1}{2r_{0}^2}\frac{\partial^{2}}{\partial\theta^{2}}=\frac{m^{2}}{2r_{0}^2}.
\end{equation}
Accordingly, for the matrix elements (\ref{eq:matrixel}) we obtain
\begin{eqnarray}
 A_{\alpha{\alpha'}} &=&
 \delta_{m,m'+1}\langle \phi_n^m|r|\phi_{n'}^{m'}\rangle 
\approx r_0\delta_{m,m'+1}\langle \phi_n^m|\phi_{n'}^{m'}\rangle \nonumber \\
&=& r_0\delta_{m,m'+1}\delta_{n,n'}.
\label{matr}
\end{eqnarray}
This allows one to restrict summations in Eqs.~(\ref{eq.g1classexpand}) 
and (\ref{eq.g2classexpand}) to just
the leading term with $n_\alpha=n_\beta=n_\gamma=0$.

Second, the fact that the energy eigenvalues $\epsilon_{n}^{m}$ separate
into $m$-independent and angular parts,
\begin{equation}
\epsilon_{n}^{m}=\epsilon_{n}+\frac{m^{2}}{2r_{0}^2},
\end{equation}
allows one to calculate the exponentially decaying terms for
$G_1(s_1,s_2)$ in Eq. (\ref{eq.g1classexpand}). More precisely, the only relevant energy differences are
\begin{equation}
\epsilon_{0}^{1}-\epsilon_{0}^{0}=\frac{1}{2r_{0}^2}=\frac{1}{2\eta},
\end{equation}
and
\begin{equation}
\epsilon_{0}^{2}-\epsilon_{0}^{1}=\frac{3}{2r_{0}^2}=\frac{3}{2{\eta}}
=3(\epsilon_{0}^{1}-\epsilon_{0}^{0}).
\label{eq.kappa2}
\end{equation}
Using Eqs.~(\ref{matr})-(\ref{eq.kappa2}), we then find that
Eq.~(\ref{eq.g1classexpand}) reduces to $G_1(s_1,s_2)=\eta e^{-|s_1-s_2|/{2{\eta}}}$.
Going back to natural units, using Eqs.~(\ref{eq:rescaling}) and~(\ref{eq:eta}), 
this gives the quasi-condensate equation of state $\rho\simeq \mu/g$, and we recover 
Eq.~(\ref{eq.g1quasicond}) of the main text.

Considering now Eq. (\ref{eq.g2classexpand}), together with the other required
cases for time ordering, a similar albeit more lengthy calculation shows that Eq.~(\ref{eq.g2classexpand}) 
reduces to Eq.~(\ref{eq.Cft}) of the main text.

\section{Ideal Bose gas limit}
\label{sec:idealgas}

The limit $\eta\ll-1$ 
corresponds to the highly degenerate ideal Bose-gas regime. In this case, the classical field problem
can be mapped onto a two-dimensional quantum harmonic oscillator. Here 
we show how the classical field approximation recovers the correlation functions expected for the ideal Bose gas.

For  $\eta\ll -1$, the Hamiltonian (\ref{eq:ham_qm}) becomes quadratic,
\begin{equation}
{ H} \simeq \frac{1}{2}\left(p_x^2+p_y^2\right)+
|\eta| \left(x^{2}+y^{2}\right),
\end{equation}
and its matrix elements can be obtained from the standard results for the quantum harmonic
oscillator with frequency $\omega = \sqrt{2|\eta|}$. We thus have
\begin{eqnarray}
  \label{eq:harm_osc}
  \langle \alpha | x |0\rangle = \langle \alpha | y |0\rangle = (2\omega)^{-1/2},
\end{eqnarray}
where $n_\alpha=0$ and $m_\alpha=1$ corresponds to the first excited state with energy 
$\epsilon_\alpha-\epsilon_0=\omega=\sqrt{2\eta}$. Then, Eq.~(\ref{eq.g1classexpand}) becomes
\begin{eqnarray}
  \label{eq:g1_harmonic}
  G_1(s_1,s_2)&=&e^{-|s_1-s_2|(\epsilon_\alpha-\epsilon_0)} |\langle \alpha|x+i
  y|0\rangle|^2 \nonumber\\&=& \frac{1}{\sqrt{2|\eta|}} e^{-\sqrt{2|\eta|}|s_1-s_2|}.
\end{eqnarray}

Going back to natural units, using
Eqs.~(\ref{eq:rescaling}) and (\ref{eq:eta}), we have 
$\rho=\psi_0^2/\sqrt{2|\eta|}= \sqrt{mk_B^2T^2/2\hbar^2|\mu|}$ and therefore
\begin{equation}
G_1(z_1,z_2)=\rho e^{-|z_1-z_2| mk_BT/\hbar^2\rho} = \rho e^{-|z_1-z_2|/l_\phi},
\end{equation}
which is the result for a highly degenerate ideal Bose gas.

The calculation of the $G_2$ function is more elaborate as there are different terms
on the right-hand side of Eq.~(\ref{eq.g2classexpand}) to compute, for
different orderings of $s_1,s_2,s_3$, and $s_4$. However, the analogy with a
simple harmonic oscillator makes the calculation possible, leading to
the recovery of the Wick's theorem (valid for quadratic Hamiltonians) and therefore
\begin{eqnarray}
G_2(s_1,s_2,s_3,s_4)&=&G_1(s_1-s_2)G_1(s_3-s_4) \nonumber \\
&+&G_1(s_1-s_4)G_1(s_2-s_3).
\end{eqnarray}
This immediately leads to the first two terms on the right-hand side of
Eq. (\ref{eq:g2free}). The last (regular) term in Eq. (\ref{eq:g2free}) is
identically zero in a noninteracting gas, whereas the third (delta-function)
term, which comes from the commutator
$[\psi(s_2),\psi^{*}(s_3)]=\delta(s_2-s_3)$, has a negligible contribution in
the highly degenerate ideal Bose-gas regime considered here.


\begin{thebibliography}{70}%
\makeatletter
\providecommand \@ifxundefined [1]{%
 \@ifx{#1\undefined}
}%
\providecommand \@ifnum [1]{%
 \ifnum #1\expandafter \@firstoftwo
 \else \expandafter \@secondoftwo
 \fi
}%
\providecommand \@ifx [1]{%
 \ifx #1\expandafter \@firstoftwo
 \else \expandafter \@secondoftwo
 \fi
}%
\providecommand \natexlab [1]{#1}%
\providecommand \enquote  [1]{``#1''}%
\providecommand \bibnamefont  [1]{#1}%
\providecommand \bibfnamefont [1]{#1}%
\providecommand \citenamefont [1]{#1}%
\providecommand \href@noop [0]{\@secondoftwo}%
\providecommand \href [0]{\begingroup \@sanitize@url \@href}%
\providecommand \@href[1]{\@@startlink{#1}\@@href}%
\providecommand \@@href[1]{\endgroup#1\@@endlink}%
\providecommand \@sanitize@url [0]{\catcode `\\12\catcode `\$12\catcode
  `\&12\catcode `\#12\catcode `\^12\catcode `\_12\catcode `\%12\relax}%
\providecommand \@@startlink[1]{}%
\providecommand \@@endlink[0]{}%
\providecommand \url  [0]{\begingroup\@sanitize@url \@url }%
\providecommand \@url [1]{\endgroup\@href {#1}{\urlprefix }}%
\providecommand \urlprefix  [0]{URL }%
\providecommand \Eprint [0]{\href }%
\providecommand \doibase [0]{http://dx.doi.org/}%
\providecommand \selectlanguage [0]{\@gobble}%
\providecommand \bibinfo  [0]{\@secondoftwo}%
\providecommand \bibfield  [0]{\@secondoftwo}%
\providecommand \translation [1]{[#1]}%
\providecommand \BibitemOpen [0]{}%
\providecommand \bibitemStop [0]{}%
\providecommand \bibitemNoStop [0]{.\EOS\space}%
\providecommand \EOS [0]{\spacefactor3000\relax}%
\providecommand \BibitemShut  [1]{\csname bibitem#1\endcsname}%
\let\auto@bib@innerbib\@empty
\bibitem [{\citenamefont {Greiner}\ \emph {et~al.}(2005)\citenamefont
  {Greiner}, \citenamefont {Regal}, \citenamefont {Stewart},\ and\
  \citenamefont {Jin}}]{Greiner2005}%
  \BibitemOpen
  \bibfield  {author} {\bibinfo {author} {\bibfnamefont {M.}~\bibnamefont
  {Greiner}}, \bibinfo {author} {\bibfnamefont {C.~A.}\ \bibnamefont {Regal}},
  \bibinfo {author} {\bibfnamefont {J.~T.}\ \bibnamefont {Stewart}}, \ and\
  \bibinfo {author} {\bibfnamefont {D.~S.}\ \bibnamefont {Jin}},\ }\href@noop
  {} {\bibfield  {journal} {\bibinfo  {journal} {Phys. Rev. Lett.}\ }\textbf
  {\bibinfo {volume} {94}},\ \bibinfo {pages} {110401} (\bibinfo {year}
  {2005})}\BibitemShut {NoStop}%
\bibitem [{\citenamefont {F\"{o}lling}\ \emph {et~al.}(2005)\citenamefont
  {F\"{o}lling}, \citenamefont {A.}, \citenamefont {Widera}, \citenamefont
  {Mandel}, \citenamefont {Gericke},\ and\ \citenamefont
  {Bloch}}]{Folling2005}%
  \BibitemOpen
  \bibfield  {author} {\bibinfo {author} {\bibfnamefont {S.}~\bibnamefont
  {F\"{o}lling}}, \bibinfo {author} {\bibfnamefont {F.~G.}\ \bibnamefont {A.}},
  \bibinfo {author} {\bibnamefont {Widera}}, \bibinfo {author} {\bibfnamefont
  {O.}~\bibnamefont {Mandel}}, \bibinfo {author} {\bibfnamefont
  {T.}~\bibnamefont {Gericke}}, \ and\ \bibinfo {author} {\bibfnamefont
  {I.}~\bibnamefont {Bloch}},\ }\href@noop {} {\bibfield  {journal} {\bibinfo
  {journal} {Nature}\ }\textbf {\bibinfo {volume} {434}},\ \bibinfo {pages}
  {481} (\bibinfo {year} {2005})}\BibitemShut {NoStop}%
\bibitem [{\citenamefont {Est\`eve}\ \emph {et~al.}(2006)\citenamefont
  {Est\`eve}, \citenamefont {Trebbia}, \citenamefont {Schumm}, \citenamefont
  {Aspect}, \citenamefont {Westbrook},\ and\ \citenamefont
  {Bouchoule}}]{Esteve06}%
  \BibitemOpen
  \bibfield  {author} {\bibinfo {author} {\bibfnamefont {J.}~\bibnamefont
  {Est\`eve}}, \bibinfo {author} {\bibfnamefont {J.-B.}\ \bibnamefont
  {Trebbia}}, \bibinfo {author} {\bibfnamefont {T.}~\bibnamefont {Schumm}},
  \bibinfo {author} {\bibfnamefont {A.}~\bibnamefont {Aspect}}, \bibinfo
  {author} {\bibfnamefont {C.~I.}\ \bibnamefont {Westbrook}}, \ and\ \bibinfo
  {author} {\bibfnamefont {I.}~\bibnamefont {Bouchoule}},\ }\href {\doibase
  10.1103/PhysRevLett.96.130403} {\bibfield  {journal} {\bibinfo  {journal}
  {Phys. Rev. Lett.}\ }\textbf {\bibinfo {volume} {96}},\ \bibinfo {eid}
  {130403} (\bibinfo {year} {2006})}\BibitemShut {NoStop}%
\bibitem [{\citenamefont {Chuu}\ \emph {et~al.}(2005)\citenamefont {Chuu},
  \citenamefont {Schreck}, \citenamefont {Meyrath}, \citenamefont {Hanssen},
  \citenamefont {Price},\ and\ \citenamefont {Raizen}}]{Raizen:2005}%
  \BibitemOpen
  \bibfield  {author} {\bibinfo {author} {\bibfnamefont {C.-S.}\ \bibnamefont
  {Chuu}}, \bibinfo {author} {\bibfnamefont {F.}~\bibnamefont {Schreck}},
  \bibinfo {author} {\bibfnamefont {T.~P.}\ \bibnamefont {Meyrath}}, \bibinfo
  {author} {\bibfnamefont {J.~L.}\ \bibnamefont {Hanssen}}, \bibinfo {author}
  {\bibfnamefont {G.~N.}\ \bibnamefont {Price}}, \ and\ \bibinfo {author}
  {\bibfnamefont {M.~G.}\ \bibnamefont {Raizen}},\ }\href {\doibase
  10.1103/PhysRevLett.95.260403} {\bibfield  {journal} {\bibinfo  {journal}
  {Phys. Rev. Lett.}\ }\textbf {\bibinfo {volume} {95}},\ \bibinfo {pages}
  {260403} (\bibinfo {year} {2005})}\BibitemShut {NoStop}%
\bibitem [{\citenamefont {Sherson}\ \emph {et~al.}(2010)\citenamefont
  {Sherson}, \citenamefont {Weitenberg}, \citenamefont {Endres}, \citenamefont
  {Cheneau}, \citenamefont {Bloch},\ and\ \citenamefont {Kuhr}}]{Sherson2010}%
  \BibitemOpen
  \bibfield  {author} {\bibinfo {author} {\bibfnamefont {J.~F.}\ \bibnamefont
  {Sherson}}, \bibinfo {author} {\bibfnamefont {C.}~\bibnamefont {Weitenberg}},
  \bibinfo {author} {\bibfnamefont {M.}~\bibnamefont {Endres}}, \bibinfo
  {author} {\bibfnamefont {M.}~\bibnamefont {Cheneau}}, \bibinfo {author}
  {\bibfnamefont {I.}~\bibnamefont {Bloch}}, \ and\ \bibinfo {author}
  {\bibfnamefont {S.}~\bibnamefont {Kuhr}},\ }\href@noop {} {\bibfield
  {journal} {\bibinfo  {journal} {Nature}\ }\textbf {\bibinfo {volume} {467}},\
  \bibinfo {pages} {68} (\bibinfo {year} {2010})}\BibitemShut {NoStop}%
\bibitem [{\citenamefont {Bakr}\ \emph {et~al.}(2010)\citenamefont {Bakr},
  \citenamefont {Peng}, \citenamefont {Tai}, \citenamefont {Ma}, \citenamefont
  {Simon}, \citenamefont {Gillen}, \citenamefont {F\"{o}lling}, \citenamefont
  {Pollet},\ and\ \citenamefont {Greiner}}]{Bakr2010}%
  \BibitemOpen
  \bibfield  {author} {\bibinfo {author} {\bibfnamefont {W.~S.}\ \bibnamefont
  {Bakr}}, \bibinfo {author} {\bibfnamefont {A.}~\bibnamefont {Peng}}, \bibinfo
  {author} {\bibfnamefont {M.~E.}\ \bibnamefont {Tai}}, \bibinfo {author}
  {\bibfnamefont {R.}~\bibnamefont {Ma}}, \bibinfo {author} {\bibfnamefont
  {J.}~\bibnamefont {Simon}}, \bibinfo {author} {\bibfnamefont {J.~I.}\
  \bibnamefont {Gillen}}, \bibinfo {author} {\bibfnamefont {S.}~\bibnamefont
  {F\"{o}lling}}, \bibinfo {author} {\bibfnamefont {L.}~\bibnamefont {Pollet}},
  \ and\ \bibinfo {author} {\bibfnamefont {M.}~\bibnamefont {Greiner}},\ }\href
  {http://www.sciencemag.org/content/329/5991/547.abstract} {\bibfield
  {journal} {\bibinfo  {journal} {Science}\ }\textbf {\bibinfo {volume}
  {329}},\ \bibinfo {pages} {547} (\bibinfo {year} {2010})}\BibitemShut
  {NoStop}%
\bibitem [{\citenamefont {\"Ottl}\ \emph {et~al.}(2005)\citenamefont {\"Ottl},
  \citenamefont {Ritter}, \citenamefont {K\"ohl},\ and\ \citenamefont
  {Esslinger}}]{Esslinger2005}%
  \BibitemOpen
  \bibfield  {author} {\bibinfo {author} {\bibfnamefont {A.}~\bibnamefont
  {\"Ottl}}, \bibinfo {author} {\bibfnamefont {S.}~\bibnamefont {Ritter}},
  \bibinfo {author} {\bibfnamefont {M.}~\bibnamefont {K\"ohl}}, \ and\ \bibinfo
  {author} {\bibfnamefont {T.}~\bibnamefont {Esslinger}},\ }\href {\doibase
  10.1103/PhysRevLett.95.090404} {\bibfield  {journal} {\bibinfo  {journal}
  {Phys. Rev. Lett.}\ }\textbf {\bibinfo {volume} {95}},\ \bibinfo {pages}
  {090404} (\bibinfo {year} {2005})}\BibitemShut {NoStop}%
\bibitem [{\citenamefont {Schellekens}\ \emph {et~al.}(2005)\citenamefont
  {Schellekens}, \citenamefont {Hoppeler}, \citenamefont {Perrin},
  \citenamefont {Gomes}, \citenamefont {Boiron}, \citenamefont {Aspect},\ and\
  \citenamefont {Westbrook}}]{HBT}%
  \BibitemOpen
  \bibfield  {author} {\bibinfo {author} {\bibfnamefont {M.}~\bibnamefont
  {Schellekens}}, \bibinfo {author} {\bibfnamefont {R.}~\bibnamefont
  {Hoppeler}}, \bibinfo {author} {\bibfnamefont {A.}~\bibnamefont {Perrin}},
  \bibinfo {author} {\bibfnamefont {J.~V.}\ \bibnamefont {Gomes}}, \bibinfo
  {author} {\bibfnamefont {D.}~\bibnamefont {Boiron}}, \bibinfo {author}
  {\bibfnamefont {A.}~\bibnamefont {Aspect}}, \ and\ \bibinfo {author}
  {\bibfnamefont {C.~I.}\ \bibnamefont {Westbrook}},\ }\href@noop {} {\bibfield
   {journal} {\bibinfo  {journal} {Science}\ } (\bibinfo {year}
  {2005})}\BibitemShut {NoStop}%
\bibitem [{\citenamefont {Jeltes}\ \emph {et~al.}(2007)\citenamefont {Jeltes},
  \citenamefont {McNamara}, \citenamefont {Hogervorst}, \citenamefont {Vassen},
  \citenamefont {Krachmalnicoff}, \citenamefont {Schellekens}, \citenamefont
  {Perrin}, \citenamefont {Chang}, \citenamefont {Boiron}, \citenamefont
  {Aspect},\ and\ \citenamefont {Westbrook}}]{Jeltes:07}%
  \BibitemOpen
  \bibfield  {author} {\bibinfo {author} {\bibfnamefont {T.}~\bibnamefont
  {Jeltes}}, \bibinfo {author} {\bibfnamefont {J.~M.}\ \bibnamefont
  {McNamara}}, \bibinfo {author} {\bibfnamefont {W.}~\bibnamefont
  {Hogervorst}}, \bibinfo {author} {\bibfnamefont {W.}~\bibnamefont {Vassen}},
  \bibinfo {author} {\bibfnamefont {V.}~\bibnamefont {Krachmalnicoff}},
  \bibinfo {author} {\bibfnamefont {M.}~\bibnamefont {Schellekens}}, \bibinfo
  {author} {\bibfnamefont {A.}~\bibnamefont {Perrin}}, \bibinfo {author}
  {\bibfnamefont {H.}~\bibnamefont {Chang}}, \bibinfo {author} {\bibfnamefont
  {D.}~\bibnamefont {Boiron}}, \bibinfo {author} {\bibfnamefont
  {A.}~\bibnamefont {Aspect}}, \ and\ \bibinfo {author} {\bibfnamefont {C.~I.}\
  \bibnamefont {Westbrook}},\ }\href@noop {} {\bibfield  {journal} {\bibinfo
  {journal} {Nature}\ }\textbf {\bibinfo {volume} {445}},\ \bibinfo {pages}
  {402} (\bibinfo {year} {2007})}\BibitemShut {NoStop}%
\bibitem [{\citenamefont {Hodgman}\ \emph
  {et~al.}(2011{\natexlab{a}})\citenamefont {Hodgman}, \citenamefont {Dall},
  \citenamefont {Manning}, \citenamefont {Baldwin},\ and\ \citenamefont
  {Truscott}}]{Hodgman2011}%
  \BibitemOpen
  \bibfield  {author} {\bibinfo {author} {\bibfnamefont {S.~S.}\ \bibnamefont
  {Hodgman}}, \bibinfo {author} {\bibfnamefont {R.~G.}\ \bibnamefont {Dall}},
  \bibinfo {author} {\bibfnamefont {A.~G.}\ \bibnamefont {Manning}}, \bibinfo
  {author} {\bibfnamefont {K.~G.~H.}\ \bibnamefont {Baldwin}}, \ and\ \bibinfo
  {author} {\bibfnamefont {A.~G.}\ \bibnamefont {Truscott}},\ }\href {\doibase
  10.1126/science.1198481} {\bibfield  {journal} {\bibinfo  {journal}
  {Science}\ }\textbf {\bibinfo {volume} {331}},\ \bibinfo {pages} {1046}
  (\bibinfo {year} {2011}{\natexlab{a}})}\BibitemShut {NoStop}%
\bibitem [{\citenamefont {Vassen}\ \emph {et~al.}(2012)\citenamefont {Vassen},
  \citenamefont {Cohen-Tannoudji}, \citenamefont {Leduc}, \citenamefont
  {Boiron}, \citenamefont {Westbrook}, \citenamefont {Truscott}, \citenamefont
  {Baldwin}, \citenamefont {Birkl}, \citenamefont {Cancio},\ and\ \citenamefont
  {Trippenbach}}]{He-Review}%
  \BibitemOpen
  \bibfield  {author} {\bibinfo {author} {\bibfnamefont {W.}~\bibnamefont
  {Vassen}}, \bibinfo {author} {\bibfnamefont {C.}~\bibnamefont
  {Cohen-Tannoudji}}, \bibinfo {author} {\bibfnamefont {M.}~\bibnamefont
  {Leduc}}, \bibinfo {author} {\bibfnamefont {D.}~\bibnamefont {Boiron}},
  \bibinfo {author} {\bibfnamefont {C.~I.}\ \bibnamefont {Westbrook}}, \bibinfo
  {author} {\bibfnamefont {A.}~\bibnamefont {Truscott}}, \bibinfo {author}
  {\bibfnamefont {K.}~\bibnamefont {Baldwin}}, \bibinfo {author} {\bibfnamefont
  {G.}~\bibnamefont {Birkl}}, \bibinfo {author} {\bibfnamefont
  {P.}~\bibnamefont {Cancio}}, \ and\ \bibinfo {author} {\bibfnamefont
  {M.}~\bibnamefont {Trippenbach}},\ }\href {\doibase
  10.1103/RevModPhys.84.175} {\bibfield  {journal} {\bibinfo  {journal} {Rev.
  Mod. Phys.}\ }\textbf {\bibinfo {volume} {84}},\ \bibinfo {pages} {175}
  (\bibinfo {year} {2012})}\BibitemShut {NoStop}%
\bibitem [{\citenamefont {Guarrera}\ \emph {et~al.}(2011)\citenamefont
  {Guarrera}, \citenamefont {W\"urtz}, \citenamefont {Ewerbeck}, \citenamefont
  {Vogler}, \citenamefont {Barontini},\ and\ \citenamefont
  {Ott}}]{SEMdetection}%
  \BibitemOpen
  \bibfield  {author} {\bibinfo {author} {\bibfnamefont {V.}~\bibnamefont
  {Guarrera}}, \bibinfo {author} {\bibfnamefont {P.}~\bibnamefont {W\"urtz}},
  \bibinfo {author} {\bibfnamefont {A.}~\bibnamefont {Ewerbeck}}, \bibinfo
  {author} {\bibfnamefont {A.}~\bibnamefont {Vogler}}, \bibinfo {author}
  {\bibfnamefont {G.}~\bibnamefont {Barontini}}, \ and\ \bibinfo {author}
  {\bibfnamefont {H.}~\bibnamefont {Ott}},\ }\href {\doibase
  10.1103/PhysRevLett.107.160403} {\bibfield  {journal} {\bibinfo  {journal}
  {Phys. Rev. Lett.}\ }\textbf {\bibinfo {volume} {107}},\ \bibinfo {pages}
  {160403} (\bibinfo {year} {2011})}\BibitemShut {NoStop}%
\bibitem [{\citenamefont {Kinoshita}\ \emph {et~al.}(2005)\citenamefont
  {Kinoshita}, \citenamefont {Wenger},\ and\ \citenamefont
  {Weiss}}]{Weiss:2005}%
  \BibitemOpen
  \bibfield  {author} {\bibinfo {author} {\bibfnamefont {T.}~\bibnamefont
  {Kinoshita}}, \bibinfo {author} {\bibfnamefont {T.}~\bibnamefont {Wenger}}, \
  and\ \bibinfo {author} {\bibfnamefont {D.~S.}\ \bibnamefont {Weiss}},\ }\href
  {\doibase 10.1103/PhysRevLett.95.190406} {\bibfield  {journal} {\bibinfo
  {journal} {Phys. Rev. Lett.}\ }\textbf {\bibinfo {volume} {95}},\ \bibinfo
  {pages} {190406} (\bibinfo {year} {2005})}\BibitemShut {NoStop}%
\bibitem [{\citenamefont {Tolra}\ \emph {et~al.}(2004)\citenamefont {Tolra},
  \citenamefont {O'Hara}, \citenamefont {Huckans}, \citenamefont {Phillips},
  \citenamefont {Rolston},\ and\ \citenamefont {Porto}}]{Phillips-2004}%
  \BibitemOpen
  \bibfield  {author} {\bibinfo {author} {\bibfnamefont {B.~L.}\ \bibnamefont
  {Tolra}}, \bibinfo {author} {\bibfnamefont {K.~M.}\ \bibnamefont {O'Hara}},
  \bibinfo {author} {\bibfnamefont {J.~H.}\ \bibnamefont {Huckans}}, \bibinfo
  {author} {\bibfnamefont {W.~D.}\ \bibnamefont {Phillips}}, \bibinfo {author}
  {\bibfnamefont {S.~L.}\ \bibnamefont {Rolston}}, \ and\ \bibinfo {author}
  {\bibfnamefont {J.~V.}\ \bibnamefont {Porto}},\ }\href {\doibase
  10.1103/PhysRevLett.92.190401} {\bibfield  {journal} {\bibinfo  {journal}
  {Phys. Rev. Lett.}\ }\textbf {\bibinfo {volume} {92}},\ \bibinfo {pages}
  {190401} (\bibinfo {year} {2004})}\BibitemShut {NoStop}%
\bibitem [{\citenamefont {Itah}\ \emph {et~al.}(2010)\citenamefont {Itah},
  \citenamefont {Veksler}, \citenamefont {Lahav}, \citenamefont {Blumkin},
  \citenamefont {Moreno}, \citenamefont {Gordon},\ and\ \citenamefont
  {Steinhauer}}]{Itah-3body}%
  \BibitemOpen
  \bibfield  {author} {\bibinfo {author} {\bibfnamefont {A.}~\bibnamefont
  {Itah}}, \bibinfo {author} {\bibfnamefont {H.}~\bibnamefont {Veksler}},
  \bibinfo {author} {\bibfnamefont {O.}~\bibnamefont {Lahav}}, \bibinfo
  {author} {\bibfnamefont {A.}~\bibnamefont {Blumkin}}, \bibinfo {author}
  {\bibfnamefont {C.}~\bibnamefont {Moreno}}, \bibinfo {author} {\bibfnamefont
  {C.}~\bibnamefont {Gordon}}, \ and\ \bibinfo {author} {\bibfnamefont
  {J.}~\bibnamefont {Steinhauer}},\ }\href {\doibase
  10.1103/PhysRevLett.104.113001} {\bibfield  {journal} {\bibinfo  {journal}
  {Phys. Rev. Lett.}\ }\textbf {\bibinfo {volume} {104}},\ \bibinfo {pages}
  {113001} (\bibinfo {year} {2010})}\BibitemShut {NoStop}%
\bibitem [{\citenamefont {Whitlock}\ \emph {et~al.}(2010)\citenamefont
  {Whitlock}, \citenamefont {Ockeloen},\ and\ \citenamefont
  {Spreeuw}}]{Whitlock:2010}%
  \BibitemOpen
  \bibfield  {author} {\bibinfo {author} {\bibfnamefont {S.}~\bibnamefont
  {Whitlock}}, \bibinfo {author} {\bibfnamefont {C.~F.}\ \bibnamefont
  {Ockeloen}}, \ and\ \bibinfo {author} {\bibfnamefont {R.~J.~C.}\ \bibnamefont
  {Spreeuw}},\ }\href {\doibase 10.1103/PhysRevLett.104.120402} {\bibfield
  {journal} {\bibinfo  {journal} {Phys. Rev. Lett.}\ }\textbf {\bibinfo
  {volume} {104}},\ \bibinfo {pages} {120402} (\bibinfo {year}
  {2010})}\BibitemShut {NoStop}%
\bibitem [{\citenamefont {T.~Rom}\ \emph {et~al.}(2006)\citenamefont {T.~Rom},
  \citenamefont {van Oosten}, \citenamefont {Schneider}, \citenamefont
  {Fölling}, \citenamefont {Paredes},\ and\ \citenamefont {Bloch}}]{Rom2006}%
  \BibitemOpen
  \bibfield  {author} {\bibinfo {author} {\bibfnamefont {T.~B.}\ \bibnamefont
  {T.~Rom}}, \bibinfo {author} {\bibfnamefont {D.}~\bibnamefont {van Oosten}},
  \bibinfo {author} {\bibfnamefont {U.}~\bibnamefont {Schneider}}, \bibinfo
  {author} {\bibfnamefont {S.}~\bibnamefont {Fölling}}, \bibinfo {author}
  {\bibfnamefont {B.}~\bibnamefont {Paredes}}, \ and\ \bibinfo {author}
  {\bibfnamefont {I.}~\bibnamefont {Bloch}},\ }\href@noop {} {\bibfield
  {journal} {\bibinfo  {journal} {Nature}\ }\textbf {\bibinfo {volume} {444}},\
  \bibinfo {pages} {733} (\bibinfo {year} {2006})}\BibitemShut {NoStop}%
\bibitem [{\citenamefont {Dall}\ \emph {et~al.}(2010)\citenamefont {Dall},
  \citenamefont {Hodgman}, \citenamefont {Manning}, \citenamefont {Johnsson},
  \citenamefont {Baldwin},\ and\ \citenamefont {Truscott}}]{Truscott:speckle}%
  \BibitemOpen
  \bibfield  {author} {\bibinfo {author} {\bibfnamefont {R.~G.}\ \bibnamefont
  {Dall}}, \bibinfo {author} {\bibfnamefont {S.~S.}\ \bibnamefont {Hodgman}},
  \bibinfo {author} {\bibfnamefont {A.~G.}\ \bibnamefont {Manning}}, \bibinfo
  {author} {\bibfnamefont {M.~T.}\ \bibnamefont {Johnsson}}, \bibinfo {author}
  {\bibfnamefont {K.~G.~H.}\ \bibnamefont {Baldwin}}, \ and\ \bibinfo {author}
  {\bibfnamefont {A.~G.}\ \bibnamefont {Truscott}},\ }\href@noop {} {\bibfield
  {journal} {\bibinfo  {journal} {Nature Communications}\ }\textbf {\bibinfo
  {volume} {2}},\ \bibinfo {pages} {291} (\bibinfo {year} {2010})}\BibitemShut
  {NoStop}%
\bibitem [{\citenamefont {Perrin}\ \emph {et~al.}(2012)\citenamefont {Perrin},
  \citenamefont {Bcker}, \citenamefont {Manz}, \citenamefont {Betz},
  \citenamefont {Koller}, \citenamefont {Plisson}, \citenamefont {Schumm},\
  and\ \citenamefont {Schmiedmayer}}]{HBTPerrin}%
  \BibitemOpen
  \bibfield  {author} {\bibinfo {author} {\bibfnamefont {A.}~\bibnamefont
  {Perrin}}, \bibinfo {author} {\bibfnamefont {R.}~\bibnamefont {Bcker}},
  \bibinfo {author} {\bibfnamefont {S.}~\bibnamefont {Manz}}, \bibinfo {author}
  {\bibfnamefont {T.}~\bibnamefont {Betz}}, \bibinfo {author} {\bibfnamefont
  {C.}~\bibnamefont {Koller}}, \bibinfo {author} {\bibfnamefont
  {T.}~\bibnamefont {Plisson}}, \bibinfo {author} {\bibfnamefont
  {T.}~\bibnamefont {Schumm}}, \ and\ \bibinfo {author} {\bibfnamefont
  {J.}~\bibnamefont {Schmiedmayer}},\ }\href@noop {} {\bibfield  {journal}
  {\bibinfo  {journal} {Nature Physics}\ } (\bibinfo {year}
  {2012})}\BibitemShut {NoStop}%
\bibitem [{\citenamefont {Hodgman}\ \emph
  {et~al.}(2011{\natexlab{b}})\citenamefont {Hodgman}, \citenamefont {Dall},
  \citenamefont {Manning}, \citenamefont {Baldwin},\ and\ \citenamefont
  {Truscott}}]{ANU-He}%
  \BibitemOpen
  \bibfield  {author} {\bibinfo {author} {\bibfnamefont {S.~S.}\ \bibnamefont
  {Hodgman}}, \bibinfo {author} {\bibfnamefont {R.~G.}\ \bibnamefont {Dall}},
  \bibinfo {author} {\bibfnamefont {A.~G.}\ \bibnamefont {Manning}}, \bibinfo
  {author} {\bibfnamefont {K.~G.~H.}\ \bibnamefont {Baldwin}}, \ and\ \bibinfo
  {author} {\bibfnamefont {A.~G.}\ \bibnamefont {Truscott}},\ }\href@noop {}
  {\bibfield  {journal} {\bibinfo  {journal} {Science}\ }\textbf {\bibinfo
  {volume} {331}},\ \bibinfo {pages} {1046} (\bibinfo {year}
  {2011}{\natexlab{b}})}\BibitemShut {NoStop}%
\bibitem [{\citenamefont {Dettmer}\ \emph {et~al.}(2001)\citenamefont
  {Dettmer}, \citenamefont {Hellweg}, \citenamefont {Ryytty}, \citenamefont
  {Arlt}, \citenamefont {Ertmer}, \citenamefont {Sengstock}, \citenamefont
  {Petrov}, \citenamefont {Shlyapnikov}, \citenamefont {Kreutzmann},
  \citenamefont {Santos},\ and\ \citenamefont
  {Lewenstein}}]{Phase-fluctuating-Hannover}%
  \BibitemOpen
  \bibfield  {author} {\bibinfo {author} {\bibfnamefont {S.}~\bibnamefont
  {Dettmer}}, \bibinfo {author} {\bibfnamefont {D.}~\bibnamefont {Hellweg}},
  \bibinfo {author} {\bibfnamefont {P.}~\bibnamefont {Ryytty}}, \bibinfo
  {author} {\bibfnamefont {J.~J.}\ \bibnamefont {Arlt}}, \bibinfo {author}
  {\bibfnamefont {W.}~\bibnamefont {Ertmer}}, \bibinfo {author} {\bibfnamefont
  {K.}~\bibnamefont {Sengstock}}, \bibinfo {author} {\bibfnamefont {D.~S.}\
  \bibnamefont {Petrov}}, \bibinfo {author} {\bibfnamefont {G.~V.}\
  \bibnamefont {Shlyapnikov}}, \bibinfo {author} {\bibfnamefont
  {H.}~\bibnamefont {Kreutzmann}}, \bibinfo {author} {\bibfnamefont
  {L.}~\bibnamefont {Santos}}, \ and\ \bibinfo {author} {\bibfnamefont
  {M.}~\bibnamefont {Lewenstein}},\ }\href {\doibase
  10.1103/PhysRevLett.87.160406} {\bibfield  {journal} {\bibinfo  {journal}
  {Phys. Rev. Lett.}\ }\textbf {\bibinfo {volume} {87}},\ \bibinfo {pages}
  {160406} (\bibinfo {year} {2001})}\BibitemShut {NoStop}%
\bibitem [{\citenamefont {Manz}\ \emph {et~al.}(2010)\citenamefont {Manz},
  \citenamefont {B\"{u}cker}, \citenamefont {Betz}, \citenamefont {Koller},
  \citenamefont {Hofferberth}, \citenamefont {Mazets}, \citenamefont
  {Imambekov}, \citenamefont {Demler}, \citenamefont {Perrin}, \citenamefont
  {Schmiedmayer},\ and\ \citenamefont {Schumm}}]{Manz2010}%
  \BibitemOpen
  \bibfield  {author} {\bibinfo {author} {\bibfnamefont {S.}~\bibnamefont
  {Manz}}, \bibinfo {author} {\bibfnamefont {R.}~\bibnamefont {B\"{u}cker}},
  \bibinfo {author} {\bibfnamefont {T.}~\bibnamefont {Betz}}, \bibinfo {author}
  {\bibfnamefont {C.}~\bibnamefont {Koller}}, \bibinfo {author} {\bibfnamefont
  {S.}~\bibnamefont {Hofferberth}}, \bibinfo {author} {\bibfnamefont {I.~E.}\
  \bibnamefont {Mazets}}, \bibinfo {author} {\bibfnamefont {A.}~\bibnamefont
  {Imambekov}}, \bibinfo {author} {\bibfnamefont {E.}~\bibnamefont {Demler}},
  \bibinfo {author} {\bibfnamefont {A.}~\bibnamefont {Perrin}}, \bibinfo
  {author} {\bibfnamefont {J.}~\bibnamefont {Schmiedmayer}}, \ and\ \bibinfo
  {author} {\bibfnamefont {T.}~\bibnamefont {Schumm}},\ }\href {\doibase
  10.1103/PhysRevA.81.031610} {\bibfield  {journal} {\bibinfo  {journal} {Phys.
  Rev. A}\ }\textbf {\bibinfo {volume} {81}},\ \bibinfo {pages} {031610}
  (\bibinfo {year} {2010})}\BibitemShut {NoStop}%
\bibitem [{\citenamefont {Spielman}\ \emph {et~al.}(2007)\citenamefont
  {Spielman}, \citenamefont {Phillips},\ and\ \citenamefont
  {Porto}}]{Spielman2007}%
  \BibitemOpen
  \bibfield  {author} {\bibinfo {author} {\bibfnamefont {I.~B.}\ \bibnamefont
  {Spielman}}, \bibinfo {author} {\bibfnamefont {W.~D.}\ \bibnamefont
  {Phillips}}, \ and\ \bibinfo {author} {\bibfnamefont {J.~V.}\ \bibnamefont
  {Porto}},\ }\href {\doibase 10.1103/PhysRevLett.98.080404} {\bibfield
  {journal} {\bibinfo  {journal} {Phys. Rev. Lett.}\ }\textbf {\bibinfo
  {volume} {98}},\ \bibinfo {pages} {080404} (\bibinfo {year}
  {2007})}\BibitemShut {NoStop}%
\bibitem [{\citenamefont {Armijo}\ \emph {et~al.}(2010)\citenamefont {Armijo},
  \citenamefont {Jacqmin}, \citenamefont {Kheruntsyan},\ and\ \citenamefont
  {Bouchoule}}]{ArmijoSkew}%
  \BibitemOpen
  \bibfield  {author} {\bibinfo {author} {\bibfnamefont {J.}~\bibnamefont
  {Armijo}}, \bibinfo {author} {\bibfnamefont {T.}~\bibnamefont {Jacqmin}},
  \bibinfo {author} {\bibfnamefont {K.~V.}\ \bibnamefont {Kheruntsyan}}, \ and\
  \bibinfo {author} {\bibfnamefont {I.}~\bibnamefont {Bouchoule}},\ }\href
  {\doibase 10.1103/PhysRevLett.105.230402} {\bibfield  {journal} {\bibinfo
  {journal} {Phys. Rev. Lett.}\ }\textbf {\bibinfo {volume} {105}},\ \bibinfo
  {pages} {230402} (\bibinfo {year} {2010})}\BibitemShut {NoStop}%
\bibitem [{\citenamefont {Sanner}\ \emph {et~al.}(2010)\citenamefont {Sanner},
  \citenamefont {Su}, \citenamefont {Keshet}, \citenamefont {Gommers},
  \citenamefont {Shin}, \citenamefont {Huang},\ and\ \citenamefont
  {Ketterle}}]{Sanner10}%
  \BibitemOpen
  \bibfield  {author} {\bibinfo {author} {\bibfnamefont {C.}~\bibnamefont
  {Sanner}}, \bibinfo {author} {\bibfnamefont {E.~J.}\ \bibnamefont {Su}},
  \bibinfo {author} {\bibfnamefont {A.}~\bibnamefont {Keshet}}, \bibinfo
  {author} {\bibfnamefont {R.}~\bibnamefont {Gommers}}, \bibinfo {author}
  {\bibfnamefont {Y.-i.}\ \bibnamefont {Shin}}, \bibinfo {author}
  {\bibfnamefont {W.}~\bibnamefont {Huang}}, \ and\ \bibinfo {author}
  {\bibfnamefont {W.}~\bibnamefont {Ketterle}},\ }\href {\doibase
  10.1103/PhysRevLett.105.040402} {\bibfield  {journal} {\bibinfo  {journal}
  {Phys. Rev. Lett.}\ }\textbf {\bibinfo {volume} {105}},\ \bibinfo {pages}
  {040402} (\bibinfo {year} {2010})}\BibitemShut {NoStop}%
\bibitem [{\citenamefont {M\"uller}\ \emph {et~al.}(2010)\citenamefont
  {M\"uller}, \citenamefont {Zimmermann}, \citenamefont {Meineke},
  \citenamefont {Brantut}, \citenamefont {Esslinger},\ and\ \citenamefont
  {Moritz}}]{Mueller10}%
  \BibitemOpen
  \bibfield  {author} {\bibinfo {author} {\bibfnamefont {T.}~\bibnamefont
  {M\"uller}}, \bibinfo {author} {\bibfnamefont {B.}~\bibnamefont
  {Zimmermann}}, \bibinfo {author} {\bibfnamefont {J.}~\bibnamefont {Meineke}},
  \bibinfo {author} {\bibfnamefont {J.-P.}\ \bibnamefont {Brantut}}, \bibinfo
  {author} {\bibfnamefont {T.}~\bibnamefont {Esslinger}}, \ and\ \bibinfo
  {author} {\bibfnamefont {H.}~\bibnamefont {Moritz}},\ }\href {\doibase
  10.1103/PhysRevLett.105.040401} {\bibfield  {journal} {\bibinfo  {journal}
  {Phys. Rev. Lett.}\ }\textbf {\bibinfo {volume} {105}},\ \bibinfo {pages}
  {040401} (\bibinfo {year} {2010})}\BibitemShut {NoStop}%
\bibitem [{\citenamefont {Sanner}\ \emph {et~al.}(2011)\citenamefont {Sanner},
  \citenamefont {Su}, \citenamefont {Keshet}, \citenamefont {Huang},
  \citenamefont {Gillen}, \citenamefont {Gommers},\ and\ \citenamefont
  {Ketterle}}]{Sanner:11}%
  \BibitemOpen
  \bibfield  {author} {\bibinfo {author} {\bibfnamefont {C.}~\bibnamefont
  {Sanner}}, \bibinfo {author} {\bibfnamefont {E.~J.}\ \bibnamefont {Su}},
  \bibinfo {author} {\bibfnamefont {A.}~\bibnamefont {Keshet}}, \bibinfo
  {author} {\bibfnamefont {W.}~\bibnamefont {Huang}}, \bibinfo {author}
  {\bibfnamefont {J.}~\bibnamefont {Gillen}}, \bibinfo {author} {\bibfnamefont
  {R.}~\bibnamefont {Gommers}}, \ and\ \bibinfo {author} {\bibfnamefont
  {W.}~\bibnamefont {Ketterle}},\ }\href {\doibase
  10.1103/PhysRevLett.106.010402} {\bibfield  {journal} {\bibinfo  {journal}
  {Phys. Rev. Lett.}\ }\textbf {\bibinfo {volume} {106}},\ \bibinfo {pages}
  {010402} (\bibinfo {year} {2011})}\BibitemShut {NoStop}%
\bibitem [{\citenamefont {Meineke}\ \emph {et~al.}(2012)\citenamefont
  {Meineke}, \citenamefont {Brantut}, \citenamefont {Stadler}, \citenamefont
  {M\"{u}ller}, \citenamefont {Moritz},\ and\ \citenamefont
  {Esslinger}}]{Esslinger2012}%
  \BibitemOpen
  \bibfield  {author} {\bibinfo {author} {\bibfnamefont {J.}~\bibnamefont
  {Meineke}}, \bibinfo {author} {\bibfnamefont {J.-P.}\ \bibnamefont
  {Brantut}}, \bibinfo {author} {\bibfnamefont {D.}~\bibnamefont {Stadler}},
  \bibinfo {author} {\bibfnamefont {T.}~\bibnamefont {M\"{u}ller}}, \bibinfo
  {author} {\bibfnamefont {H.}~\bibnamefont {Moritz}}, \ and\ \bibinfo {author}
  {\bibfnamefont {T.}~\bibnamefont {Esslinger}},\ }\href@noop {} {\bibfield
  {journal} {\bibinfo  {journal} {1202.5250}\ } (\bibinfo {year}
  {2012})}\BibitemShut {NoStop}%
\bibitem [{\citenamefont {Hung}\ \emph
  {et~al.}(2011{\natexlab{a}})\citenamefont {Hung}, \citenamefont {Zhang},
  \citenamefont {Gemelke},\ and\ \citenamefont {Chin}}]{Hung2011}%
  \BibitemOpen
  \bibfield  {author} {\bibinfo {author} {\bibfnamefont {C.-H.}\ \bibnamefont
  {Hung}}, \bibinfo {author} {\bibfnamefont {X.}~\bibnamefont {Zhang}},
  \bibinfo {author} {\bibfnamefont {N.}~\bibnamefont {Gemelke}}, \ and\
  \bibinfo {author} {\bibfnamefont {C.}~\bibnamefont {Chin}},\ }\href@noop {}
  {\bibfield  {journal} {\bibinfo  {journal} {Nature}\ }\textbf {\bibinfo
  {volume} {470}},\ \bibinfo {pages} {236} (\bibinfo {year}
  {2011}{\natexlab{a}})}\BibitemShut {NoStop}%
\bibitem [{\citenamefont {Armijo}\ \emph {et~al.}(2011)\citenamefont {Armijo},
  \citenamefont {Jacqmin}, \citenamefont {Kheruntsyan},\ and\ \citenamefont
  {Bouchoule}}]{Armijo10_2}%
  \BibitemOpen
  \bibfield  {author} {\bibinfo {author} {\bibfnamefont {J.}~\bibnamefont
  {Armijo}}, \bibinfo {author} {\bibfnamefont {T.}~\bibnamefont {Jacqmin}},
  \bibinfo {author} {\bibfnamefont {K.}~\bibnamefont {Kheruntsyan}}, \ and\
  \bibinfo {author} {\bibfnamefont {I.}~\bibnamefont {Bouchoule}},\ }\href
  {\doibase 10.1103/PhysRevA.83.021605} {\bibfield  {journal} {\bibinfo
  {journal} {Phys. Rev. A}\ }\textbf {\bibinfo {volume} {83}},\ \bibinfo
  {pages} {021605} (\bibinfo {year} {2011})}\BibitemShut {NoStop}%
\bibitem [{\citenamefont {Jacqmin}\ \emph {et~al.}(2011)\citenamefont
  {Jacqmin}, \citenamefont {Armijo}, \citenamefont {Berrada}, \citenamefont
  {Kheruntsyan},\ and\ \citenamefont {Bouchoule}}]{Subbunching}%
  \BibitemOpen
  \bibfield  {author} {\bibinfo {author} {\bibfnamefont {T.}~\bibnamefont
  {Jacqmin}}, \bibinfo {author} {\bibfnamefont {J.}~\bibnamefont {Armijo}},
  \bibinfo {author} {\bibfnamefont {T.}~\bibnamefont {Berrada}}, \bibinfo
  {author} {\bibfnamefont {K.~V.}\ \bibnamefont {Kheruntsyan}}, \ and\ \bibinfo
  {author} {\bibfnamefont {I.}~\bibnamefont {Bouchoule}},\ }\href {\doibase
  10.1103/PhysRevLett.106.230405} {\bibfield  {journal} {\bibinfo  {journal}
  {Phys. Rev. Lett.}\ }\textbf {\bibinfo {volume} {106}},\ \bibinfo {pages}
  {230405} (\bibinfo {year} {2011})}\BibitemShut {NoStop}%
\bibitem [{\citenamefont {Est\`eve}\ \emph {et~al.}(2008)\citenamefont
  {Est\`eve}, \citenamefont {Gross}, \citenamefont {Weller}, \citenamefont
  {Giovanazzi},\ and\ \citenamefont {Oberthaler}}]{Esteve2008}%
  \BibitemOpen
  \bibfield  {author} {\bibinfo {author} {\bibfnamefont {J.}~\bibnamefont
  {Est\`eve}}, \bibinfo {author} {\bibfnamefont {C.}~\bibnamefont {Gross}},
  \bibinfo {author} {\bibfnamefont {A.}~\bibnamefont {Weller}}, \bibinfo
  {author} {\bibfnamefont {S.}~\bibnamefont {Giovanazzi}}, \ and\ \bibinfo
  {author} {\bibfnamefont {M.~K.}\ \bibnamefont {Oberthaler}},\ }\href@noop {}
  {\bibfield  {journal} {\bibinfo  {journal} {Nature}\ }\textbf {\bibinfo
  {volume} {455}},\ \bibinfo {pages} {1216} (\bibinfo {year}
  {2008})}\BibitemShut {NoStop}%
\bibitem [{\citenamefont {Gross}\ \emph {et~al.}(2010)\citenamefont {Gross},
  \citenamefont {Zibold}, \citenamefont {Nicklas}, \citenamefont {Est\`eve},\
  and\ \citenamefont {Oberthaler}}]{Oberthaler-interferometer}%
  \BibitemOpen
  \bibfield  {author} {\bibinfo {author} {\bibfnamefont {C.}~\bibnamefont
  {Gross}}, \bibinfo {author} {\bibfnamefont {T.}~\bibnamefont {Zibold}},
  \bibinfo {author} {\bibfnamefont {E.}~\bibnamefont {Nicklas}}, \bibinfo
  {author} {\bibfnamefont {J.}~\bibnamefont {Est\`eve}}, \ and\ \bibinfo
  {author} {\bibfnamefont {M.~K.}\ \bibnamefont {Oberthaler}},\ }\href@noop {}
  {\bibfield  {journal} {\bibinfo  {journal} {{Nature}}\ }\textbf {\bibinfo
  {volume} {464}},\ \bibinfo {pages} {1165} (\bibinfo {year}
  {2010})}\BibitemShut {NoStop}%
\bibitem [{\citenamefont {Riedel}\ \emph {et~al.}(2010)\citenamefont {Riedel},
  \citenamefont {B\"ohi}, \citenamefont {Li}, \citenamefont {H\"ansch},
  \citenamefont {Sinatra},\ and\ \citenamefont
  {Treutlein}}]{Riedel-interferometer}%
  \BibitemOpen
  \bibfield  {author} {\bibinfo {author} {\bibfnamefont {M.~F.}\ \bibnamefont
  {Riedel}}, \bibinfo {author} {\bibfnamefont {P.}~\bibnamefont {B\"ohi}},
  \bibinfo {author} {\bibfnamefont {Y.}~\bibnamefont {Li}}, \bibinfo {author}
  {\bibfnamefont {T.~W.}\ \bibnamefont {H\"ansch}}, \bibinfo {author}
  {\bibfnamefont {A.}~\bibnamefont {Sinatra}}, \ and\ \bibinfo {author}
  {\bibfnamefont {P.}~\bibnamefont {Treutlein}},\ }\href@noop {} {\bibfield
  {journal} {\bibinfo  {journal} {{Nature}}\ }\textbf {\bibinfo {volume}
  {464}},\ \bibinfo {pages} {1170} (\bibinfo {year} {2010})}\BibitemShut
  {NoStop}%
\bibitem [{\citenamefont {L\"ucke}\ \emph {et~al.}(2011)\citenamefont
  {L\"ucke}, \citenamefont {Scherer}, \citenamefont {Kruse}, \citenamefont
  {Pezz\'e}, \citenamefont {Deuretzbacher}, \citenamefont {Hyllus},
  \citenamefont {Topic}, \citenamefont {Peise}, \citenamefont {Ertmer},
  \citenamefont {Arlt}, \citenamefont {Santos}, \citenamefont {Smerzi},\ and\
  \citenamefont {Klempt}}]{Hannover-twins}%
  \BibitemOpen
  \bibfield  {author} {\bibinfo {author} {\bibfnamefont {B.}~\bibnamefont
  {L\"ucke}}, \bibinfo {author} {\bibfnamefont {M.}~\bibnamefont {Scherer}},
  \bibinfo {author} {\bibfnamefont {J.}~\bibnamefont {Kruse}}, \bibinfo
  {author} {\bibfnamefont {L.}~\bibnamefont {Pezz\'e}}, \bibinfo {author}
  {\bibfnamefont {F.}~\bibnamefont {Deuretzbacher}}, \bibinfo {author}
  {\bibfnamefont {P.}~\bibnamefont {Hyllus}}, \bibinfo {author} {\bibfnamefont
  {O.}~\bibnamefont {Topic}}, \bibinfo {author} {\bibfnamefont
  {J.}~\bibnamefont {Peise}}, \bibinfo {author} {\bibfnamefont
  {W.}~\bibnamefont {Ertmer}}, \bibinfo {author} {\bibfnamefont
  {J.}~\bibnamefont {Arlt}}, \bibinfo {author} {\bibfnamefont {L.}~\bibnamefont
  {Santos}}, \bibinfo {author} {\bibfnamefont {A.}~\bibnamefont {Smerzi}}, \
  and\ \bibinfo {author} {\bibfnamefont {C.}~\bibnamefont {Klempt}},\
  }\href@noop {} {\bibfield  {journal} {\bibinfo  {journal} {Science}\ }\textbf
  {\bibinfo {volume} {334}},\ \bibinfo {pages} {773} (\bibinfo {year}
  {2011})}\BibitemShut {NoStop}%
\bibitem [{\citenamefont {Bookjans}\ \emph {et~al.}(2011)\citenamefont
  {Bookjans}, \citenamefont {Hamley},\ and\ \citenamefont
  {Chapman}}]{Chapman:11}%
  \BibitemOpen
  \bibfield  {author} {\bibinfo {author} {\bibfnamefont {E.~M.}\ \bibnamefont
  {Bookjans}}, \bibinfo {author} {\bibfnamefont {C.~D.}\ \bibnamefont
  {Hamley}}, \ and\ \bibinfo {author} {\bibfnamefont {M.~S.}\ \bibnamefont
  {Chapman}},\ }\href {\doibase 10.1103/PhysRevLett.107.210406} {\bibfield
  {journal} {\bibinfo  {journal} {Phys. Rev. Lett.}\ }\textbf {\bibinfo
  {volume} {107}},\ \bibinfo {pages} {210406} (\bibinfo {year}
  {2011})}\BibitemShut {NoStop}%
\bibitem [{\citenamefont {Jaskula}\ \emph {et~al.}(2010)\citenamefont
  {Jaskula}, \citenamefont {Bonneau}, \citenamefont {Partridge}, \citenamefont
  {Krachmalnicoff}, \citenamefont {Deuar}, \citenamefont {Kheruntsyan},
  \citenamefont {Aspect}, \citenamefont {Boiron},\ and\ \citenamefont
  {Westbrook}}]{Jaskula2010}%
  \BibitemOpen
  \bibfield  {author} {\bibinfo {author} {\bibfnamefont {J.-C.}\ \bibnamefont
  {Jaskula}}, \bibinfo {author} {\bibfnamefont {M.}~\bibnamefont {Bonneau}},
  \bibinfo {author} {\bibfnamefont {G.~B.}\ \bibnamefont {Partridge}}, \bibinfo
  {author} {\bibfnamefont {V.}~\bibnamefont {Krachmalnicoff}}, \bibinfo
  {author} {\bibfnamefont {P.}~\bibnamefont {Deuar}}, \bibinfo {author}
  {\bibfnamefont {K.~V.}\ \bibnamefont {Kheruntsyan}}, \bibinfo {author}
  {\bibfnamefont {A.}~\bibnamefont {Aspect}}, \bibinfo {author} {\bibfnamefont
  {D.}~\bibnamefont {Boiron}}, \ and\ \bibinfo {author} {\bibfnamefont {C.~I.}\
  \bibnamefont {Westbrook}},\ }\href {\doibase 10.1103/PhysRevLett.105.190402}
  {\bibfield  {journal} {\bibinfo  {journal} {Phys. Rev. Lett.}\ }\textbf
  {\bibinfo {volume} {105}},\ \bibinfo {pages} {190402} (\bibinfo {year}
  {2010})}\BibitemShut {NoStop}%
\bibitem [{\citenamefont {B\"ucker}\ \emph {et~al.}(2011)\citenamefont
  {B\"ucker}, \citenamefont {Grond}, \citenamefont {Manz}, \citenamefont
  {Berrada}, \citenamefont {Betz}, \citenamefont {Koller}, \citenamefont
  {Hohenester}, \citenamefont {Schumm}, \citenamefont {Perrin},\ and\
  \citenamefont {Schmiedmayer}}]{Vienna-twins}%
  \BibitemOpen
  \bibfield  {author} {\bibinfo {author} {\bibfnamefont {R.}~\bibnamefont
  {B\"ucker}}, \bibinfo {author} {\bibfnamefont {J.}~\bibnamefont {Grond}},
  \bibinfo {author} {\bibfnamefont {S.}~\bibnamefont {Manz}}, \bibinfo {author}
  {\bibfnamefont {T.}~\bibnamefont {Berrada}}, \bibinfo {author} {\bibfnamefont
  {T.}~\bibnamefont {Betz}}, \bibinfo {author} {\bibfnamefont {C.}~\bibnamefont
  {Koller}}, \bibinfo {author} {\bibfnamefont {U.}~\bibnamefont {Hohenester}},
  \bibinfo {author} {\bibfnamefont {T.}~\bibnamefont {Schumm}}, \bibinfo
  {author} {\bibfnamefont {A.}~\bibnamefont {Perrin}}, \ and\ \bibinfo {author}
  {\bibfnamefont {J.}~\bibnamefont {Schmiedmayer}},\ }\href@noop {} {\bibfield
  {journal} {\bibinfo  {journal} {Nature Physics}\ }\textbf {\bibinfo {volume}
  {7}},\ \bibinfo {pages} {608} (\bibinfo {year} {2011})}\BibitemShut {NoStop}%
\bibitem [{\citenamefont {Kheruntsyan}\ \emph {et~al.}(2012)\citenamefont
  {Kheruntsyan}, \citenamefont {Jaskula}, \citenamefont {Deuar}, \citenamefont
  {Bonneau}, \citenamefont {Partridge}, \citenamefont {Ruaudel}, \citenamefont
  {Lopes}, \citenamefont {Boiron},\ and\ \citenamefont
  {Westbrook}}]{Kheruntsyan-CS}%
  \BibitemOpen
  \bibfield  {author} {\bibinfo {author} {\bibfnamefont {K.~V.}\ \bibnamefont
  {Kheruntsyan}}, \bibinfo {author} {\bibfnamefont {J.-C.}\ \bibnamefont
  {Jaskula}}, \bibinfo {author} {\bibfnamefont {P.}~\bibnamefont {Deuar}},
  \bibinfo {author} {\bibfnamefont {M.}~\bibnamefont {Bonneau}}, \bibinfo
  {author} {\bibfnamefont {G.~B.}\ \bibnamefont {Partridge}}, \bibinfo {author}
  {\bibfnamefont {J.}~\bibnamefont {Ruaudel}}, \bibinfo {author} {\bibfnamefont
  {R.}~\bibnamefont {Lopes}}, \bibinfo {author} {\bibfnamefont
  {D.}~\bibnamefont {Boiron}}, \ and\ \bibinfo {author} {\bibfnamefont {C.~I.}\
  \bibnamefont {Westbrook}},\ }\href {\doibase 10.1103/PhysRevLett.108.260401}
  {\bibfield  {journal} {\bibinfo  {journal} {Phys. Rev. Lett.}\ }\textbf
  {\bibinfo {volume} {108}},\ \bibinfo {pages} {260401} (\bibinfo {year}
  {2012})}\BibitemShut {NoStop}%
\bibitem [{\citenamefont {Altman}\ \emph {et~al.}(2004)\citenamefont {Altman},
  \citenamefont {Demler},\ and\ \citenamefont {Lukin}}]{Altman04}%
  \BibitemOpen
  \bibfield  {author} {\bibinfo {author} {\bibfnamefont {E.}~\bibnamefont
  {Altman}}, \bibinfo {author} {\bibfnamefont {E.}~\bibnamefont {Demler}}, \
  and\ \bibinfo {author} {\bibfnamefont {M.~D.}\ \bibnamefont {Lukin}},\ }\href
  {\doibase 10.1103/PhysRevA.70.013603} {\bibfield  {journal} {\bibinfo
  {journal} {Phys. Rev. A}\ }\textbf {\bibinfo {volume} {70}},\ \bibinfo
  {pages} {013603} (\bibinfo {year} {2004})}\BibitemShut {NoStop}%
\bibitem [{\citenamefont {Shvarchuck}\ \emph {et~al.}(2002)\citenamefont
  {Shvarchuck}, \citenamefont {Buggle}, \citenamefont {Petrov}, \citenamefont
  {Dieckmann}, \citenamefont {Zielonkowski}, \citenamefont {Kemmann},
  \citenamefont {Tiecke}, \citenamefont {von Klitzing}, \citenamefont
  {Shlyapnikov},\ and\ \citenamefont {Walraven}}]{Shvarchuck2002}%
  \BibitemOpen
  \bibfield  {author} {\bibinfo {author} {\bibfnamefont {I.}~\bibnamefont
  {Shvarchuck}}, \bibinfo {author} {\bibfnamefont {C.}~\bibnamefont {Buggle}},
  \bibinfo {author} {\bibfnamefont {D.~S.}\ \bibnamefont {Petrov}}, \bibinfo
  {author} {\bibfnamefont {K.}~\bibnamefont {Dieckmann}}, \bibinfo {author}
  {\bibfnamefont {M.}~\bibnamefont {Zielonkowski}}, \bibinfo {author}
  {\bibfnamefont {M.}~\bibnamefont {Kemmann}}, \bibinfo {author} {\bibfnamefont
  {T.~G.}\ \bibnamefont {Tiecke}}, \bibinfo {author} {\bibfnamefont
  {W.}~\bibnamefont {von Klitzing}}, \bibinfo {author} {\bibfnamefont {G.~V.}\
  \bibnamefont {Shlyapnikov}}, \ and\ \bibinfo {author} {\bibfnamefont
  {J.~T.~M.}\ \bibnamefont {Walraven}},\ }\href {\doibase
  10.1103/PhysRevLett.89.270404} {\bibfield  {journal} {\bibinfo  {journal}
  {Phys. Rev. Lett.}\ }\textbf {\bibinfo {volume} {89}},\ \bibinfo {pages}
  {270404} (\bibinfo {year} {2002})}\BibitemShut {NoStop}%
\bibitem [{\citenamefont {van Amerongen}\ \emph {et~al.}(2008)\citenamefont
  {van Amerongen}, \citenamefont {van Es}, \citenamefont {Wicke}, \citenamefont
  {Kheruntsyan},\ and\ \citenamefont {van Druten}}]{Amerongen08}%
  \BibitemOpen
  \bibfield  {author} {\bibinfo {author} {\bibfnamefont {A.~H.}\ \bibnamefont
  {van Amerongen}}, \bibinfo {author} {\bibfnamefont {J.~J.~P.}\ \bibnamefont
  {van Es}}, \bibinfo {author} {\bibfnamefont {P.}~\bibnamefont {Wicke}},
  \bibinfo {author} {\bibfnamefont {K.~V.}\ \bibnamefont {Kheruntsyan}}, \ and\
  \bibinfo {author} {\bibfnamefont {N.~J.}\ \bibnamefont {van Druten}},\ }\href
  {\doibase 10.1103/PhysRevLett.100.090402} {\bibfield  {journal} {\bibinfo
  {journal} {Phys. Rev. Lett.}\ }\textbf {\bibinfo {volume} {100}},\ \bibinfo
  {eid} {090402} (\bibinfo {year} {2008})}\BibitemShut {NoStop}%
\bibitem [{\citenamefont {Jacqmin}\ \emph {et~al.}(2012)\citenamefont
  {Jacqmin}, \citenamefont {Fang}, \citenamefont {Berrada}, \citenamefont
  {Roscilde},\ and\ \citenamefont {Bouchoule}}]{momentumdistriJacqmin}%
  \BibitemOpen
  \bibfield  {author} {\bibinfo {author} {\bibfnamefont {T.}~\bibnamefont
  {Jacqmin}}, \bibinfo {author} {\bibfnamefont {B.}~\bibnamefont {Fang}},
  \bibinfo {author} {\bibfnamefont {T.}~\bibnamefont {Berrada}}, \bibinfo
  {author} {\bibfnamefont {T.}~\bibnamefont {Roscilde}}, \ and\ \bibinfo
  {author} {\bibfnamefont {I.}~\bibnamefont {Bouchoule}},\ }\href@noop {}
  {\bibfield  {journal} {\bibinfo  {journal} {arXiv:1207.2855}\ } (\bibinfo
  {year} {2012})}\BibitemShut {NoStop}%
\bibitem [{\citenamefont {Davis}\ \emph {et~al.}(2012)\citenamefont {Davis},
  \citenamefont {Blakie}, \citenamefont {van Amerongen}, \citenamefont {van
  Druten},\ and\ \citenamefont {Kheruntsyan}}]{Davis:12}%
  \BibitemOpen
  \bibfield  {author} {\bibinfo {author} {\bibfnamefont {M.~J.}\ \bibnamefont
  {Davis}}, \bibinfo {author} {\bibfnamefont {P.~B.}\ \bibnamefont {Blakie}},
  \bibinfo {author} {\bibfnamefont {A.~H.}\ \bibnamefont {van Amerongen}},
  \bibinfo {author} {\bibfnamefont {N.~J.}\ \bibnamefont {van Druten}}, \ and\
  \bibinfo {author} {\bibfnamefont {K.~V.}\ \bibnamefont {Kheruntsyan}},\
  }\href {\doibase 10.1103/PhysRevA.85.031604} {\bibfield  {journal} {\bibinfo
  {journal} {Phys. Rev. A}\ }\textbf {\bibinfo {volume} {85}},\ \bibinfo
  {pages} {031604} (\bibinfo {year} {2012})}\BibitemShut {NoStop}%
\bibitem [{\citenamefont {Tung}\ \emph {et~al.}(2010)\citenamefont {Tung},
  \citenamefont {Lamporesi}, \citenamefont {Lobser}, \citenamefont {Xia},\ and\
  \citenamefont {Cornell}}]{2Dfocusing}%
  \BibitemOpen
  \bibfield  {author} {\bibinfo {author} {\bibfnamefont {S.}~\bibnamefont
  {Tung}}, \bibinfo {author} {\bibfnamefont {G.}~\bibnamefont {Lamporesi}},
  \bibinfo {author} {\bibfnamefont {D.}~\bibnamefont {Lobser}}, \bibinfo
  {author} {\bibfnamefont {L.}~\bibnamefont {Xia}}, \ and\ \bibinfo {author}
  {\bibfnamefont {E.~A.}\ \bibnamefont {Cornell}},\ }\href {\doibase
  10.1103/PhysRevLett.105.230408} {\bibfield  {journal} {\bibinfo  {journal}
  {Phys. Rev. Lett.}\ }\textbf {\bibinfo {volume} {105}},\ \bibinfo {pages}
  {230408} (\bibinfo {year} {2010})}\BibitemShut {NoStop}%
\bibitem [{\citenamefont {Mathey}\ \emph {et~al.}(2009)\citenamefont {Mathey},
  \citenamefont {Vishwanath},\ and\ \citenamefont {Altman}}]{Mathey2009}%
  \BibitemOpen
  \bibfield  {author} {\bibinfo {author} {\bibfnamefont {L.}~\bibnamefont
  {Mathey}}, \bibinfo {author} {\bibfnamefont {A.}~\bibnamefont {Vishwanath}},
  \ and\ \bibinfo {author} {\bibfnamefont {E.}~\bibnamefont {Altman}},\ }\href
  {\doibase 10.1103/PhysRevA.79.013609} {\bibfield  {journal} {\bibinfo
  {journal} {Phys. Rev. A}\ }\textbf {\bibinfo {volume} {79}},\ \bibinfo
  {pages} {013609} (\bibinfo {year} {2009})}\BibitemShut {NoStop}%
\bibitem [{\citenamefont {Imambekov}\ \emph {et~al.}(2009)\citenamefont
  {Imambekov}, \citenamefont {Mazets}, \citenamefont {Petrov}, \citenamefont
  {Gritsev}, \citenamefont {Manz}, \citenamefont {Hofferberth}, \citenamefont
  {Schumm}, \citenamefont {Demler},\ and\ \citenamefont
  {Schmiedmayer}}]{Imambekov09}%
  \BibitemOpen
  \bibfield  {author} {\bibinfo {author} {\bibfnamefont {A.}~\bibnamefont
  {Imambekov}}, \bibinfo {author} {\bibfnamefont {I.~E.}\ \bibnamefont
  {Mazets}}, \bibinfo {author} {\bibfnamefont {D.~S.}\ \bibnamefont {Petrov}},
  \bibinfo {author} {\bibfnamefont {V.}~\bibnamefont {Gritsev}}, \bibinfo
  {author} {\bibfnamefont {S.}~\bibnamefont {Manz}}, \bibinfo {author}
  {\bibfnamefont {S.}~\bibnamefont {Hofferberth}}, \bibinfo {author}
  {\bibfnamefont {T.}~\bibnamefont {Schumm}}, \bibinfo {author} {\bibfnamefont
  {E.}~\bibnamefont {Demler}}, \ and\ \bibinfo {author} {\bibfnamefont
  {J.}~\bibnamefont {Schmiedmayer}},\ }\href {\doibase
  10.1103/PhysRevA.80.033604} {\bibfield  {journal} {\bibinfo  {journal} {Phys.
  Rev. A}\ }\textbf {\bibinfo {volume} {80}},\ \bibinfo {pages} {033604}
  (\bibinfo {year} {2009})}\BibitemShut {NoStop}%
\bibitem [{\citenamefont {Hohenberg}(1967)}]{Hohenberg67}%
  \BibitemOpen
  \bibfield  {author} {\bibinfo {author} {\bibfnamefont {P.~C.}\ \bibnamefont
  {Hohenberg}},\ }\href {\doibase 10.1103/PhysRev.158.383} {\bibfield
  {journal} {\bibinfo  {journal} {Phys. Rev.}\ }\textbf {\bibinfo {volume}
  {158}},\ \bibinfo {pages} {383} (\bibinfo {year} {1967})}\BibitemShut
  {NoStop}%
\bibitem [{\citenamefont {Petrov}\ \emph {et~al.}(2000)\citenamefont {Petrov},
  \citenamefont {Shlyapnikov},\ and\ \citenamefont {Walraven}}]{Petrov00}%
  \BibitemOpen
  \bibfield  {author} {\bibinfo {author} {\bibfnamefont {D.~S.}\ \bibnamefont
  {Petrov}}, \bibinfo {author} {\bibfnamefont {G.~V.}\ \bibnamefont
  {Shlyapnikov}}, \ and\ \bibinfo {author} {\bibfnamefont {J.~T.~M.}\
  \bibnamefont {Walraven}},\ }\href {\doibase 10.1103/PhysRevLett.85.3745}
  {\bibfield  {journal} {\bibinfo  {journal} {Phys. Rev. Lett.}\ }\textbf
  {\bibinfo {volume} {85}},\ \bibinfo {pages} {3745} (\bibinfo {year}
  {2000})}\BibitemShut {NoStop}%
\bibitem [{\citenamefont {Popov}(1980)}]{Popov1980}%
  \BibitemOpen
  \bibfield  {author} {\bibinfo {author} {\bibfnamefont {V.~N.}\ \bibnamefont
  {Popov}},\ }\href@noop {} {\bibfield  {journal} {\bibinfo  {journal} {JETP
  Lett.}\ }\textbf {\bibinfo {volume} {31}},\ \bibinfo {pages} {526} (\bibinfo
  {year} {1980})}\BibitemShut {NoStop}%
\bibitem [{\citenamefont {Mora}\ and\ \citenamefont {Castin}(2003)}]{Mora03}%
  \BibitemOpen
  \bibfield  {author} {\bibinfo {author} {\bibfnamefont {C.}~\bibnamefont
  {Mora}}\ and\ \bibinfo {author} {\bibfnamefont {Y.}~\bibnamefont {Castin}},\
  }\href {\doibase 10.1103/PhysRevA.67.053615} {\bibfield  {journal} {\bibinfo
  {journal} {Phys. Rev. A}\ }\textbf {\bibinfo {volume} {67}},\ \bibinfo
  {pages} {053615} (\bibinfo {year} {2003})}\BibitemShut {NoStop}%
\bibitem [{\citenamefont {Cazalilla}(2004)}]{Cazalilla2004}%
  \BibitemOpen
  \bibfield  {author} {\bibinfo {author} {\bibfnamefont {M.~A.}\ \bibnamefont
  {Cazalilla}},\ }\href@noop {} {\bibfield  {journal} {\bibinfo  {journal} {J.
  Phys. B: At. Mol. Opt. Phys.}\ }\textbf {\bibinfo {volume} {37}},\ \bibinfo
  {pages} {S1} (\bibinfo {year} {2004})}\BibitemShut {NoStop}%
\bibitem [{\citenamefont {Castin}\ \emph {et~al.}(2000)\citenamefont {Castin},
  \citenamefont {Dum}, \citenamefont {Mandonnet}, \citenamefont {Minguzzi},\
  and\ \citenamefont {Carusotto}}]{Castinatomlaser}%
  \BibitemOpen
  \bibfield  {author} {\bibinfo {author} {\bibfnamefont {Y.}~\bibnamefont
  {Castin}}, \bibinfo {author} {\bibfnamefont {R.}~\bibnamefont {Dum}},
  \bibinfo {author} {\bibfnamefont {E.}~\bibnamefont {Mandonnet}}, \bibinfo
  {author} {\bibfnamefont {A.}~\bibnamefont {Minguzzi}}, \ and\ \bibinfo
  {author} {\bibfnamefont {I.}~\bibnamefont {Carusotto}},\ }\href@noop {}
  {\bibfield  {journal} {\bibinfo  {journal} {Journal of Modern Optics}\
  }\textbf {\bibinfo {volume} {47}},\ \bibinfo {pages} {2671} (\bibinfo {year}
  {2000})}\BibitemShut {NoStop}%
\bibitem [{c_f()}]{c_field_and_Bog}%
  \BibitemOpen
  \href@noop {} {}\bibinfo {note} {The results that we obtain here, such as
  Eqs. (\ref{eq.corrhw}) and (\ref{eq:pkscaling}), can be derived directly
  within the Bgoliubov approach for each domain (i.e. without invoking the
  additional classical field approximation), assuming $\langle
  \hat{n}_{k}\rangle\gg 1$.}\BibitemShut {Stop}%
\bibitem [{\citenamefont {Kheruntsyan}\ \emph {et~al.}(2003)\citenamefont
  {Kheruntsyan}, \citenamefont {Gangardt}, \citenamefont {Drummond},\ and\
  \citenamefont {Shlyapnikov}}]{KheruntsyanPRL03}%
  \BibitemOpen
  \bibfield  {author} {\bibinfo {author} {\bibfnamefont {K.~V.}\ \bibnamefont
  {Kheruntsyan}}, \bibinfo {author} {\bibfnamefont {D.~M.}\ \bibnamefont
  {Gangardt}}, \bibinfo {author} {\bibfnamefont {P.~D.}\ \bibnamefont
  {Drummond}}, \ and\ \bibinfo {author} {\bibfnamefont {G.~V.}\ \bibnamefont
  {Shlyapnikov}},\ }\href {\doibase 10.1103/PhysRevLett.91.040403} {\bibfield
  {journal} {\bibinfo  {journal} {Phys. Rev. Lett.}\ }\textbf {\bibinfo
  {volume} {91}},\ \bibinfo {pages} {040403} (\bibinfo {year}
  {2003})}\BibitemShut {NoStop}%
\bibitem [{\citenamefont {Haldane}(1981)}]{Haldane:81}%
  \BibitemOpen
  \bibfield  {author} {\bibinfo {author} {\bibfnamefont {F.~D.~M.}\
  \bibnamefont {Haldane}},\ }\href {\doibase 10.1103/PhysRevLett.47.1840}
  {\bibfield  {journal} {\bibinfo  {journal} {Phys. Rev. Lett.}\ }\textbf
  {\bibinfo {volume} {47}},\ \bibinfo {pages} {1840} (\bibinfo {year}
  {1981})}\BibitemShut {NoStop}%
\bibitem [{\citenamefont {Sykes}\ \emph {et~al.}(2008)\citenamefont {Sykes},
  \citenamefont {Gangardt}, \citenamefont {Davis}, \citenamefont {Viering},
  \citenamefont {Raizen},\ and\ \citenamefont {Kheruntsyan}}]{Sykes:08}%
  \BibitemOpen
  \bibfield  {author} {\bibinfo {author} {\bibfnamefont {A.~G.}\ \bibnamefont
  {Sykes}}, \bibinfo {author} {\bibfnamefont {D.~M.}\ \bibnamefont {Gangardt}},
  \bibinfo {author} {\bibfnamefont {M.~J.}\ \bibnamefont {Davis}}, \bibinfo
  {author} {\bibfnamefont {K.}~\bibnamefont {Viering}}, \bibinfo {author}
  {\bibfnamefont {M.~G.}\ \bibnamefont {Raizen}}, \ and\ \bibinfo {author}
  {\bibfnamefont {K.~V.}\ \bibnamefont {Kheruntsyan}},\ }\href {\doibase
  10.1103/PhysRevLett.100.160406} {\bibfield  {journal} {\bibinfo  {journal}
  {Phys. Rev. Lett.}\ }\textbf {\bibinfo {volume} {100}},\ \bibinfo {pages}
  {160406} (\bibinfo {year} {2008})}\BibitemShut {NoStop}%
\bibitem [{\citenamefont {Deuar}\ \emph {et~al.}(2009)\citenamefont {Deuar},
  \citenamefont {Sykes}, \citenamefont {Gangardt}, \citenamefont {Davis},
  \citenamefont {Drummond},\ and\ \citenamefont {Kheruntsyan}}]{Deuar:09}%
  \BibitemOpen
  \bibfield  {author} {\bibinfo {author} {\bibfnamefont {P.}~\bibnamefont
  {Deuar}}, \bibinfo {author} {\bibfnamefont {A.~G.}\ \bibnamefont {Sykes}},
  \bibinfo {author} {\bibfnamefont {D.~M.}\ \bibnamefont {Gangardt}}, \bibinfo
  {author} {\bibfnamefont {M.~J.}\ \bibnamefont {Davis}}, \bibinfo {author}
  {\bibfnamefont {P.~D.}\ \bibnamefont {Drummond}}, \ and\ \bibinfo {author}
  {\bibfnamefont {K.~V.}\ \bibnamefont {Kheruntsyan}},\ }\href {\doibase
  10.1103/PhysRevA.79.043619} {\bibfield  {journal} {\bibinfo  {journal} {Phys.
  Rev. A}\ }\textbf {\bibinfo {volume} {79}},\ \bibinfo {pages} {043619}
  (\bibinfo {year} {2009})}\BibitemShut {NoStop}%
\bibitem [{qua()}]{quantum_fluctuations}%
  \BibitemOpen
  \href@noop {} {}\bibinfo {note} {Vacuum fluctuations have a non-negligible
  effect on the correlation function $G_1(z_1,z_2)$ if the characteristic
  distance $l_T$, over which $G_1(z_1,z_2)$ decays algebraically, is much
  larger than $l_{\phi}^{(0)}$, i.e., if $l_T\gg l_{\phi}^{(0)}$. This
  condition can be rewritten as $k_BT\ll g\rho \exp(-2\pi/\sqrt{\gamma})$,
  implying that the vacuum fluctuations are important only at temperatures that
  are exponentially small in the weakly interacting regime $\gamma\ll 1$. Such
  temperatures are beyond the reach of current ultracold atom
  experiments.}\BibitemShut {Stop}%
\bibitem [{nk()}]{nk}%
  \BibitemOpen
  \href@noop {} {}\bibinfo {note} {The shot-noise term $\langle
  \hat{n}_{k}\rangle$ cannot be obtained in this approach, but it is
  negligible compared to $\langle \hat{n}_{k}\rangle ^{2} $ as long as $k\ll
  \max \{k_BT/\hbar \sqrt{\rho g/m}, \sqrt{m k_BT}/\hbar\}$. The latter
  condition is fulfilled for $k \lesssim 1/l_{\phi}$.}\BibitemShut {Stop}%
\bibitem [{lim()}]{limits}%
  \BibitemOpen
  \href@noop {} {}\bibinfo {note} {In obtaining this equation, we have
  approximated the discrete sums over $q=2l_{\phi}k$ and $q'=2l_{\phi}k'$ by
  continuous integrals, and extended the integration limits from
  ($-q_{\max},q_{\max}$) to ($-\infty,\infty$). This can be done because the
  maximum momentum cutoff $k_{\max}\sim 1/\xi$ in the Luttinger liquid theory
  \cite{Cazalilla2004} leads to $q_{\max}\sim l_{\phi}/\xi$, which is much
  larger than unity according to Eq.~(\ref{eq.Tco}), and because the function
  $\mathcal{F}(q,q')$ decays to zero sufficiently fast.}\BibitemShut {Stop}%
\bibitem [{sum()}]{sum-rule-consistency}%
  \BibitemOpen
  \href@noop {} {}\bibinfo {note} {For a more accurate and consistent estimate
  of the breakdown of the sum rule as $T$ increases and approaches
  $T_{\rm{co}}$, one would need to evaluate the contribution of $\sum _k
  \langle \hat{n}_k\rangle^2$ to the right-hand side of Eq. (30) to leading order in
  $T/T_{\rm{co}}$, which is, however, beyond the scope of this
  paper.}\BibitemShut {Stop}%
\bibitem [{reg()}]{regimes}%
  \BibitemOpen
  \href@noop {} {}\bibinfo {note} {The condition $g\rho
  e^{-2\pi/\sqrt{\gamma}}<k_BT<\sqrt{\gamma}\hbar^2\rho^2/m$ (see also
  \cite{quantum_fluctuations}) follows from the fact that at these temperatures,
  the $G_1(z_1,z_2)$ function decays exponentially and as long as the
  (small-momentum) bulk of the momentum distribution is concerned, the mode
  occupancies here are indeed high. If, on the other hand, one is concerned
  with position-space correlations \cite{Sykes:08,Deuar:09}, then the (low)
  occupancies of large-momentum modes become relevant too, in which case the
  vacuum fluctuations become relevant as soon as $g\rho>k_B T$. Accordingly, 
  for position-space correlations,
  the range of validity of the classical field theory in the quasicondensate
  regime is $g\rho<k_BT<\sqrt{\gamma}\hbar^2\rho^2/m$.}\BibitemShut {Stop}%
\bibitem [{\citenamefont {Bouchoule}\ \emph {et~al.}(2007)\citenamefont
  {Bouchoule}, \citenamefont {Kheruntsyan},\ and\ \citenamefont
  {Shlyapnikov}}]{Bouchoule07}%
  \BibitemOpen
  \bibfield  {author} {\bibinfo {author} {\bibfnamefont {I.}~\bibnamefont
  {Bouchoule}}, \bibinfo {author} {\bibfnamefont {K.~V.}\ \bibnamefont
  {Kheruntsyan}}, \ and\ \bibinfo {author} {\bibfnamefont {G.~V.}\ \bibnamefont
  {Shlyapnikov}},\ }\href {\doibase 10.1103/PhysRevA.75.031606} {\bibfield
  {journal} {\bibinfo  {journal} {Phys. Rev. A}\ }\textbf {\bibinfo {volume}
  {75}},\ \bibinfo {pages} {031606} (\bibinfo {year} {2007})}\BibitemShut
  {NoStop}%
\bibitem [{\citenamefont {Hung}\ \emph
  {et~al.}(2011{\natexlab{b}})\citenamefont {Hung}, \citenamefont {Zhang},
  \citenamefont {Gemelke},\ and\ \citenamefont {Chin}}]{ScaleInvariance2}%
  \BibitemOpen
  \bibfield  {author} {\bibinfo {author} {\bibfnamefont {C.-H.}\ \bibnamefont
  {Hung}}, \bibinfo {author} {\bibfnamefont {X.}~\bibnamefont {Zhang}},
  \bibinfo {author} {\bibfnamefont {N.}~\bibnamefont {Gemelke}}, \ and\
  \bibinfo {author} {\bibfnamefont {C.}~\bibnamefont {Chin}},\ }\href@noop {}
  {\bibfield  {journal} {\bibinfo  {journal} {Nature}\ }\textbf {\bibinfo
  {volume} {470}},\ \bibinfo {pages} {236} (\bibinfo {year}
  {2011}{\natexlab{b}})}\BibitemShut {NoStop}%
\bibitem [{\citenamefont {Wright}\ \emph {et~al.}(2012)\citenamefont {Wright},
  \citenamefont {Perrin}, \citenamefont {Bray}, \citenamefont {Schmiedmayer},\
  and\ \citenamefont {Kheruntsyan}}]{IBG}%
  \BibitemOpen
  \bibfield  {author} {\bibinfo {author} {\bibfnamefont {T.~M.}\ \bibnamefont
  {Wright}}, \bibinfo {author} {\bibfnamefont {A.}~\bibnamefont {Perrin}},
  \bibinfo {author} {\bibfnamefont {A.}~\bibnamefont {Bray}}, \bibinfo {author}
  {\bibfnamefont {J.}~\bibnamefont {Schmiedmayer}}, \ and\ \bibinfo {author}
  {\bibfnamefont {K.~V.}\ \bibnamefont {Kheruntsyan}},\ }\href {\doibase
  10.1103/PhysRevA.86.023618} {\bibfield  {journal} {\bibinfo  {journal} {Phys.
  Rev. A}\ }\textbf {\bibinfo {volume} {86}},\ \bibinfo {pages} {023618}
  (\bibinfo {year} {2012})}\BibitemShut {NoStop}%
\bibitem [{\citenamefont {Gangardt}\ and\ \citenamefont
  {Pustilnik}(2008)}]{Dima:08}%
  \BibitemOpen
  \bibfield  {author} {\bibinfo {author} {\bibfnamefont {D.~M.}\ \bibnamefont
  {Gangardt}}\ and\ \bibinfo {author} {\bibfnamefont {M.}~\bibnamefont
  {Pustilnik}},\ }\href {\doibase 10.1103/PhysRevA.77.041604} {\bibfield
  {journal} {\bibinfo  {journal} {Phys. Rev. A}\ }\textbf {\bibinfo {volume}
  {77}},\ \bibinfo {pages} {041604} (\bibinfo {year} {2008})}\BibitemShut
  {NoStop}%
\bibitem [{\citenamefont {Muth}\ \emph {et~al.}(2010)\citenamefont {Muth},
  \citenamefont {Schmidt},\ and\ \citenamefont {Fleischhauer}}]{Muth:10}%
  \BibitemOpen
  \bibfield  {author} {\bibinfo {author} {\bibfnamefont {D.}~\bibnamefont
  {Muth}}, \bibinfo {author} {\bibfnamefont {B.}~\bibnamefont {Schmidt}}, \
  and\ \bibinfo {author} {\bibfnamefont {M.}~\bibnamefont {Fleischhauer}},\
  }\href {http://stacks.iop.org/1367-2630/12/i=8/a=083065} {\bibfield
  {journal} {\bibinfo  {journal} {New Journal of Physics}\ }\textbf {\bibinfo
  {volume} {12}},\ \bibinfo {pages} {083065} (\bibinfo {year}
  {2010})}\BibitemShut {NoStop}%
\bibitem [{\citenamefont {Giraud}\ and\ \citenamefont
  {Serreau}(2010)}]{Giraud:thermalization}%
  \BibitemOpen
  \bibfield  {author} {\bibinfo {author} {\bibfnamefont {A.}~\bibnamefont
  {Giraud}}\ and\ \bibinfo {author} {\bibfnamefont {J.}~\bibnamefont
  {Serreau}},\ }\href {\doibase 10.1103/PhysRevLett.104.230405} {\bibfield
  {journal} {\bibinfo  {journal} {Phys. Rev. Lett.}\ }\textbf {\bibinfo
  {volume} {104}},\ \bibinfo {pages} {230405} (\bibinfo {year}
  {2010})}\BibitemShut {NoStop}%
\bibitem [{\citenamefont {Polkovnikov}\ \emph {et~al.}(2011)\citenamefont
  {Polkovnikov}, \citenamefont {Sengupta}, \citenamefont {Silva},\ and\
  \citenamefont {Vengalattore}}]{Polkovnikov:review}%
  \BibitemOpen
  \bibfield  {author} {\bibinfo {author} {\bibfnamefont {A.}~\bibnamefont
  {Polkovnikov}}, \bibinfo {author} {\bibfnamefont {K.}~\bibnamefont
  {Sengupta}}, \bibinfo {author} {\bibfnamefont {A.}~\bibnamefont {Silva}}, \
  and\ \bibinfo {author} {\bibfnamefont {M.}~\bibnamefont {Vengalattore}},\
  }\href {\doibase 10.1103/RevModPhys.83.863} {\bibfield  {journal} {\bibinfo
  {journal} {Rev. Mod. Phys.}\ }\textbf {\bibinfo {volume} {83}},\ \bibinfo
  {pages} {863} (\bibinfo {year} {2011})}\BibitemShut {NoStop}%
\end{thebibliography}

%

\end{document}